\documentclass[]{spie}  %>>> use for US letter paper
\usepackage{amsmath,amsfonts,amssymb,aas_macros}
\usepackage{graphicx}
\usepackage[colorlinks=true, allcolors=blue]{hyperref}
\usepackage{rotating}
\usepackage{tikz}

\title{Design and characterization of the Cosmology Large Angular Scale Surveyor (CLASS) 93 GHz focal plane}

\author[a]{Sumit Dahal}
\author[a,b]{Aamir Ali}
\author[a]{John W. Appel}
\author[c]{Thomas Essinger-Hileman}
\author[a]{Charles Bennett}
\author[a]{Michael Brewer}
\author[d]{Ricardo Bustos}
\author[a]{Manwei Chan}
\author[e]{David T. Chuss}
\author[a]{Joseph Cleary}
\author[c]{Felipe Colazo}
\author[a]{Jullianna Couto}
\author[c]{Kevin Denis}
\author[f]{Rolando D\"{u}nner}
\author[a]{Joseph Eimer}
\author[a]{Trevor Engelhoven}
\author[f]{Pedro Fluxa}
\author[h]{Mark Halpern}
\author[a]{Kathleen Harrington}
\author[c]{Kyle Helson}
\author[g]{Gene Hilton}
\author[h]{Gary Hinshaw}
\author[g]{Johannes Hubmayr}
\author[a]{Jeffrey Iuliano}
\author[a]{John Karakla}
\author[a]{Tobias Marriage}
\author[k]{Jeffrey McMahon}
\author[a]{Nathan Miller}
\author[a]{Carolina Nu\~{n}ez}
\author[a]{Ivan Padilla}
\author[i]{Gonzalo Palma}
\author[a,j]{Lucas Parker}
\author[a]{Matthew Petroff}
\author[i]{Bastian Pradenas}
\author[l]{Rodrigo Reeves}
\author[g]{Carl Reintsema}
\author[c]{Karwan Rostem}
\author[c]{Marco Sagliocca}
\author[c]{Kongpop U-Yen}
\author[a]{Deniz Valle}
\author[a]{Bingjie Wang}
\author[a]{Qinan Wang}
\author[a]{Duncan Watts}
\author[a]{Janet Weiland}
\author[c]{Edward Wollack}
\author[a,m]{Zhilei Xu}
\author[h]{Ziang Yan}
\author[n]{Lingzhen Zeng}

\affil[a]{Department of Physics and Astronomy, Johns Hopkins University, Baltimore, MD 21218, USA}
\affil[b]{Department of Physics, University of California, Berkeley, CA 94720, USA }
\affil[c]{NASA Goddard Space Flight Center, Greenbelt, MD 20771, USA }
\affil[d]{Facultad de Ingenier\'{i}a, Universidad Cat\'{o}lica de la Sant\'{i}sima Concepci\'{o}n, Alonso de Ribera 2850, Concepci\'{o}n, Chile }
\affil[e]{Department of Physics, Villanova University, Villanova, PA 19085, USA }
\affil[f]{
Instituto de Astrof\'{i}sica and Centro de Astro-Ingenier\'{i}a, Facultad de F\'{i}sica, Pontificia Universidad Cat\'{o}lica de Chile, 7820436
Macul, Santiago, Chile }
\affil[g]{National Institute of Standards and Technology, Boulder, CO 80305, USA}
\affil[h]{Department of Physics and Astronomy, University of British Columbia, Vancouver, BC V6T 1z4 ,Canada}
\affil[i]{Departmento de F\'{i}sica, FCFM, Universidad de Chile, Blanco Encalada 2008, Santiago, Chile}
\affil[j]{Space and Remote Sensing, MS D436, Los Alamos National Laboratory,
Los Alamos, NM 87544, USA}
\affil[k]{Department of Physics, University of Michigan, Ann Arbor, MI, 48109, USA}
\affil[l]{Departamento de Astronom\'{i}a, Universidad de Concepci\'{o}n, Casilla 160 C, Concepci\'{o}n, Chile}
\affil[m]{Department of Physics and Astronomy, University of Pennsylvania, Philadelphia, PA 19104, USA}
\affil[n]{Harvard-Smithsonian Center for Astrophysics, Cambridge, MA 02138, USA }

\authorinfo{Further author information: (Send correspondence to S. Dahal \& A. Ali)\\S. Dahal: E-mail: sumit.dahal@jhu.edu \\ A. Ali: Email: aali18@jhu.edu}

\begin{document} 
\maketitle
\begin{abstract}
The Cosmology Large Angular Scale Surveyor (CLASS) aims to detect and characterize the primordial B-mode signal and make a sample-variance-limited measurement of the optical depth to reionization. CLASS is a ground-based, multi-frequency microwave polarimeter that surveys 70\% of the microwave sky every day from the Atacama Desert. The focal plane detector arrays of all CLASS telescopes contain smooth-walled feedhorns that couple to transition-edge sensor (TES) bolometers through symmetric planar orthomode transducer (OMT) antennas. These low noise polarization-sensitive detector arrays are fabricated on mono-crystalline silicon wafers to maintain TES uniformity and optimize optical efficiency throughout the wafer. In this paper, we discuss the design and characterization of the first CLASS 93 GHz detector array. We measure the dark parameters, bandpass, and noise spectra of the detectors and report that the detectors are photon-noise limited. With current array yield of 82\%, we estimate the total array noise-equivalent power (NEP) to be 2.1 aW$\sqrt[]{\mathrm{s}}$.     
 
\end{abstract}

% Include a list of keywords after the abstract 

\section{INTRODUCTION}
\label{sec:intro}  % \label{} allows reference to this section
As the universe expanded and cooled from an extremely dense and hot state, free electrons combined with protons, allowing photons to decouple from matter and free stream throughout the universe. Today, we observe these relic photons as a 2.726 K \cite{fixsen} blackbody across the sky, known as the cosmic microwave background (CMB). Temperature fluctuations in the CMB have now been measured to the cosmic variance limit and have been pivotal in placing constraints on parameters of $\Lambda$CDM cosmology \cite{wmap,planck}. However, the near isotropy and homogeneity of the CMB, even over the regions of the sky that were not in causal contact at decoupling, presents a significant challenge in our understanding of cosmology. An inflationary phase at the beginning of the universe could solve this problem, as the exponential expansion would smooth out inhomogeneities to scales much larger than the observable universe \cite{guth,linde}. This inflationary paradigm can be tested and constrained by characterizing the predicted stochastic background of primordial gravitational waves through its signature in the polarization of the CMB.

CMB polarization arises from Thomson scattering of photons off free electrons situated in a temperature quadrupole. Instead of characterizing the CMB polarization anisotropy with the Stokes parameters (which depend on an arbitrary choice of coordinates), we can decompose it into non-local divergence and curl components, referred to as E-modes and B-modes respectively. Inflation predicts both scalar (over and under densities) and tensor perturbations (gravitational waves) to the space-time metric that lead to the anisotropies observed both in the temperature and polarization of the CMB. However, the scalar perturbations can only produce E-modes, while tensor perturbations can produce both E- and B-modes \cite{marc,zaldarriaga}. Therefore, a detection of primordial B-modes would be an evidence of inflation. Detecting this signal, however, is challenging. It is orders of magnitude fainter than the CMB temperature fluctuations, and Galactic synchrotron and dust emission also produce a large polarized signal at millimeter wavelengths. Given the low signal amplitude and large foreground contributions, it is essential that an experiment aiming to detect primordial B-modes pursue multi-band observations over large angular scales with large arrays of efficient background-limited detectors \cite{watts15}.

The Cosmology Large Angular Scale Surveyor (CLASS) is designed to achieve this goal by measuring or placing an upper limit on the amplitude of primordial gravitational waves parameterized by the ratio of tensor-to-scalar fluctuations (r) at a level of r = 0.01 \cite{tom14, katie}. In addition, through precise measurement of E-mode polarization at large angular scales, CLASS can make a sample variance limited measurement of the optical depth to reionization, $\tau$, \cite{watts18} which is currently the least-constrained fundamental $\Lambda$CDM parameter \cite{wmap,planck}. CLASS is a ground-based multi-frequency microwave polarimeter array that surveys 70\% of the microwave sky every day from the Atacama Desert, inside the Parque Astron\'{o}mico de Atacama \cite{ricardo}. CLASS consists of two 93 GHz telescopes optimized for CMB observation near the minimum of polarized Galactic emission, a 38 GHz telescope probing the polarized synchrotron emission, and a 145/217 GHz dichroic receiver mapping the polarized dust. These central frequencies and their band edges were chosen not only to straddle the Galactic-foreground minimum, but also to avoid strong atmospheric emission lines, primarily from oxygen and water. All CLASS telescopes use a variable-delay polarization modulator (VPM) \cite{nathan,katie18} as their first optical element in order to put the signal band at 10 Hz, well above the 1/$f$ noise that comes from a combination of instrumental and atmospheric drifts. The unique combination of large sky coverage, broad frequency range, rapid front-end polarization modulator, and background-limited detectors provides CLASS with the high sensitivity, stability, and control over systematics necessary to characterize both the reionization and recombination peaks of the CMB E- and B-mode power spectra. For further description about the CLASS instruments refer to Refs. \citenum{tom14}, \citenum{katie}, and \citenum{joseph}.

In this paper, we report on the design and characterization of the first CLASS 93 GHz (hereafter referred to as ``W-band'') focal plane detector array. See Table \ref{tab:W-survey} for a summary of the CLASS W-band Telescope characteristics. The second W-band focal plane array has the same design and is expected to have similar detector properties. The first W-band receiver was recently deployed at the CLASS site. The CLASS 38 GHz (Q-band) receiver has been operational since May 8, 2016. For more detailed descriptions on the Q-band detector array see Ref. \citenum{john14}. In Section \ref{sec:fp}, we will discuss the design of the W-band focal planes.  Section \ref{sec:det} contains the in-lab characterization of detectors in the first W-band focal plane and  estimation of array sensitivity with the observed detector parameters. We summarize the current state and future of the CLASS focal plane detector arrays in Section \ref{sec:conclusion}.

\begin{table}
\caption{CLASS W-band Telescope characteristics} 
\label{tab:W-survey}
\begin{center}       
\begin{tabular}{cc} %% this creates two columns; use |c|c| for column dividers; and add \hline after each row for row dividers
\hline
\rule[-1ex]{0pt}{3.5ex}  Frequency & 77 GHz to 108 GHz  \\
\rule[-1ex]{0pt}{3.5ex}  Beam FWHM & $40^{\prime}$  \\
\rule[-1ex]{0pt}{3.5ex}  Field of view & $\sim$ 380 sq. degrees  \\
\rule[-1ex]{0pt}{3.5ex}  Detectors & Feedhorn-coupled Transition-Edge Sensor (TES) bolometers \cite{kevin,karwan16} \\
\rule[-1ex]{0pt}{3.5ex}  Number of detectors & 518 (259 dual-polarization-sensitive pixels) \\
\rule[-1ex]{0pt}{3.5ex}  Cryogenics & Pulse-tube cooled dilution fridge with $T_\mathrm{min}$ = 26 mK \cite{jeff} \\
\rule[-1ex]{0pt}{3.5ex}  Polarization Modulation & Variable-delay Polarization Modulator (VPM) \\
\hline
\end{tabular}
\end{center}
\end{table}

\section{W-band Focal Plane Design}
\label{sec:fp}

\begin{figure}[ht]
   \begin{center}
   \begin{tabular}{c}
   \includegraphics[scale=0.5]{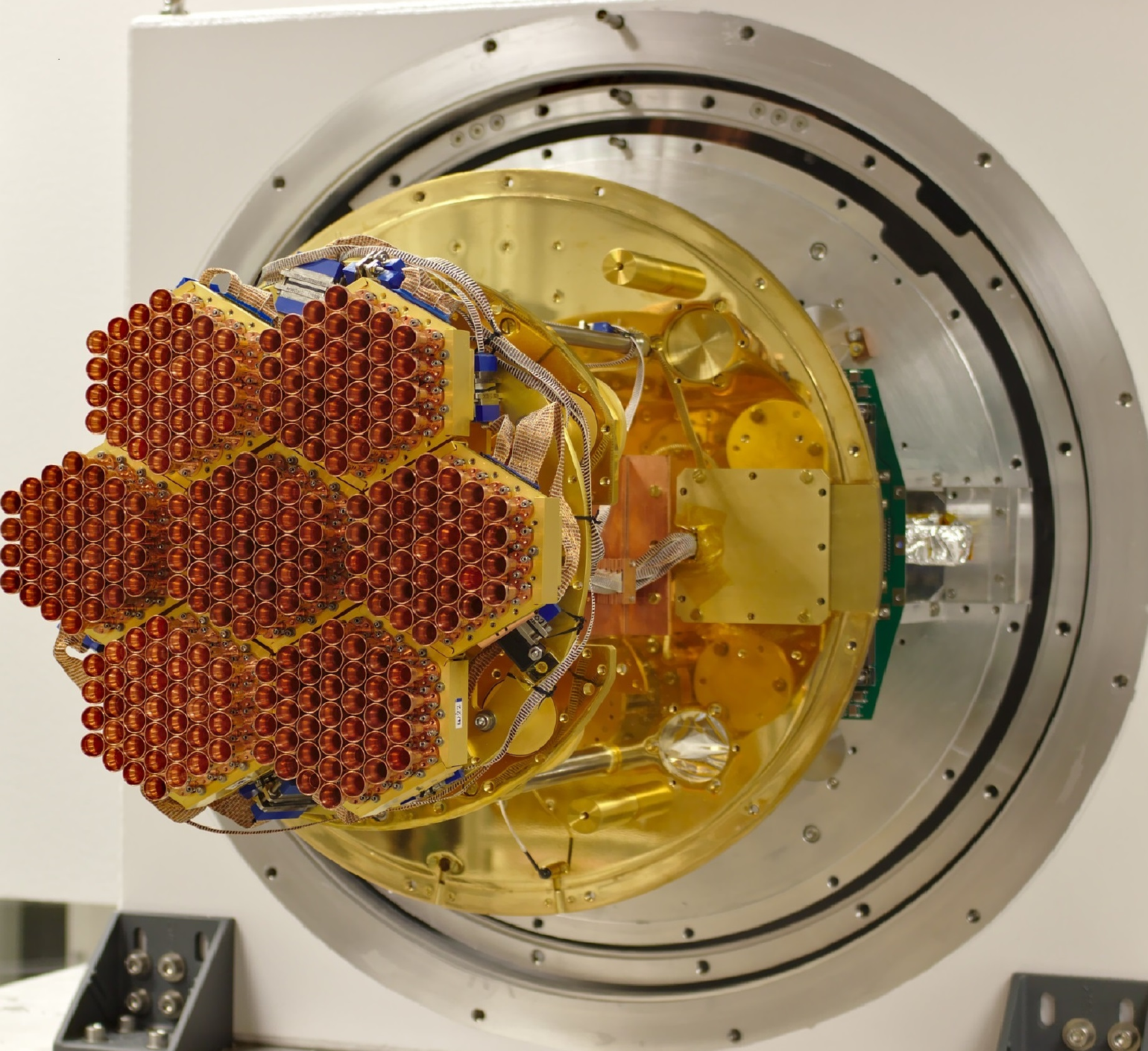}
   \end{tabular}
   \end{center}
   \caption[example] 
%>>>> use \label inside caption to get Fig. number with \ref{}
   { \label{fig:fp} Fully assembled first CLASS W-band focal plane mounted in the cryostat. The focal plane consists of seven individual detector modules mounted on a Au-plated copper web interface, which is then mounted onto the mixing chamber plate of a pulse-tube cooled dilution refrigerator. Each module contains 37 smooth-walled copper feedhorns that guide light to the dual-polarization-sensitive detectors on the focal plane.}
   \end{figure}

To achieve the required sensitivity for CLASS, the detectors have to be cooled to a stable low bath temperature. We therefore mount the CLASS W-band focal plane on the mixing chamber plate of a pulse-tube cooled dilution refrigerator with gold-plated copper posts for strong thermal contact. In the field, the focal plane is operated at a stable bath temperature of $\sim$ 35 mK \cite{jeff}. Figure \ref{fig:fp} shows the fully assembled first CLASS W-band focal plane mounted in the cryostat. The focal plane has a hexagonal modular design consisting of seven individual modules. The modular design makes it straightforward to swap modules during testing and assembly.

  \begin{sidewaysfigure}
   \begin{center}
   \begin{tabular}{cc}
   \includegraphics[scale=0.4]{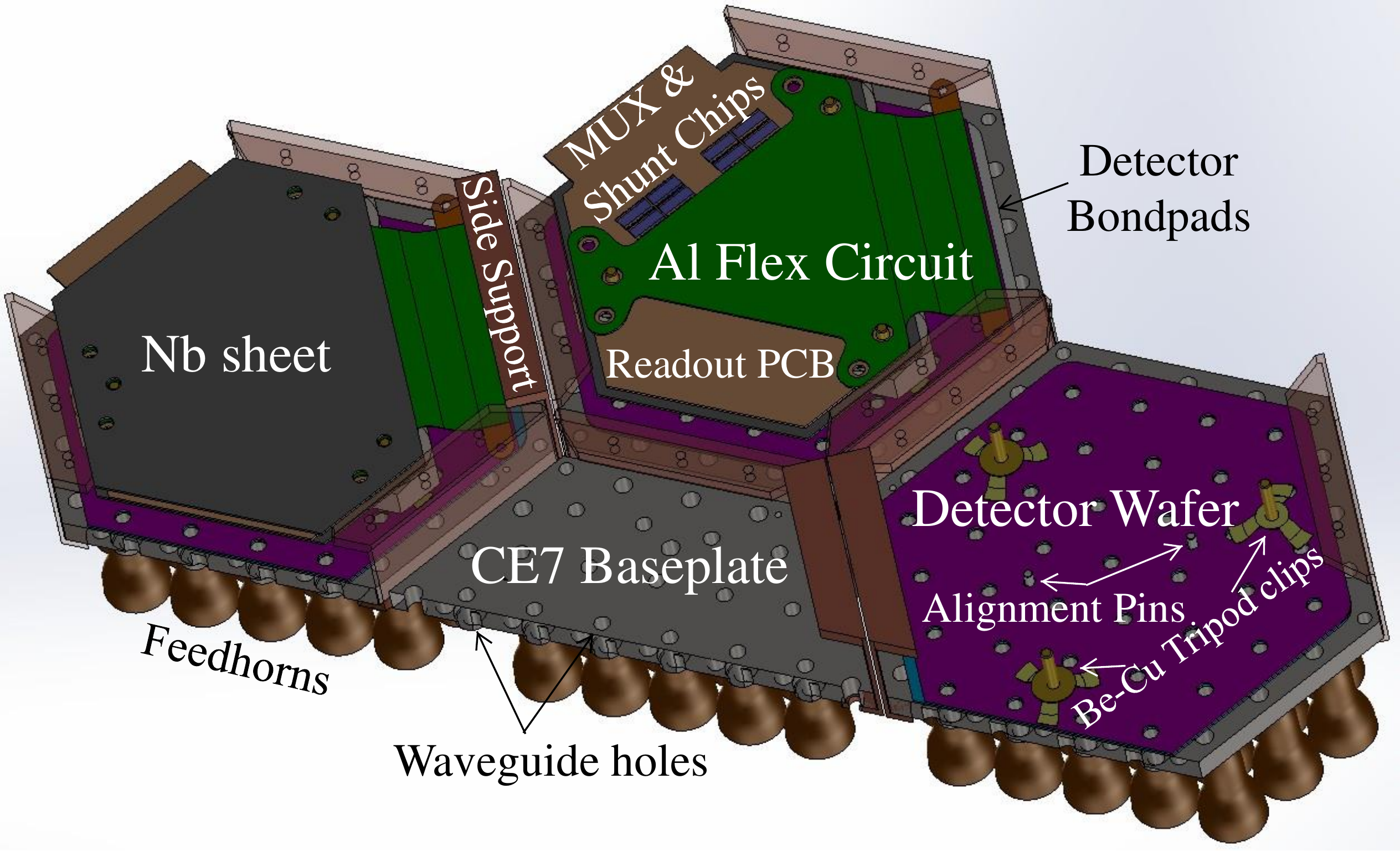}
   \includegraphics[trim={0 4.5cm 0 0},clip,scale=0.39]{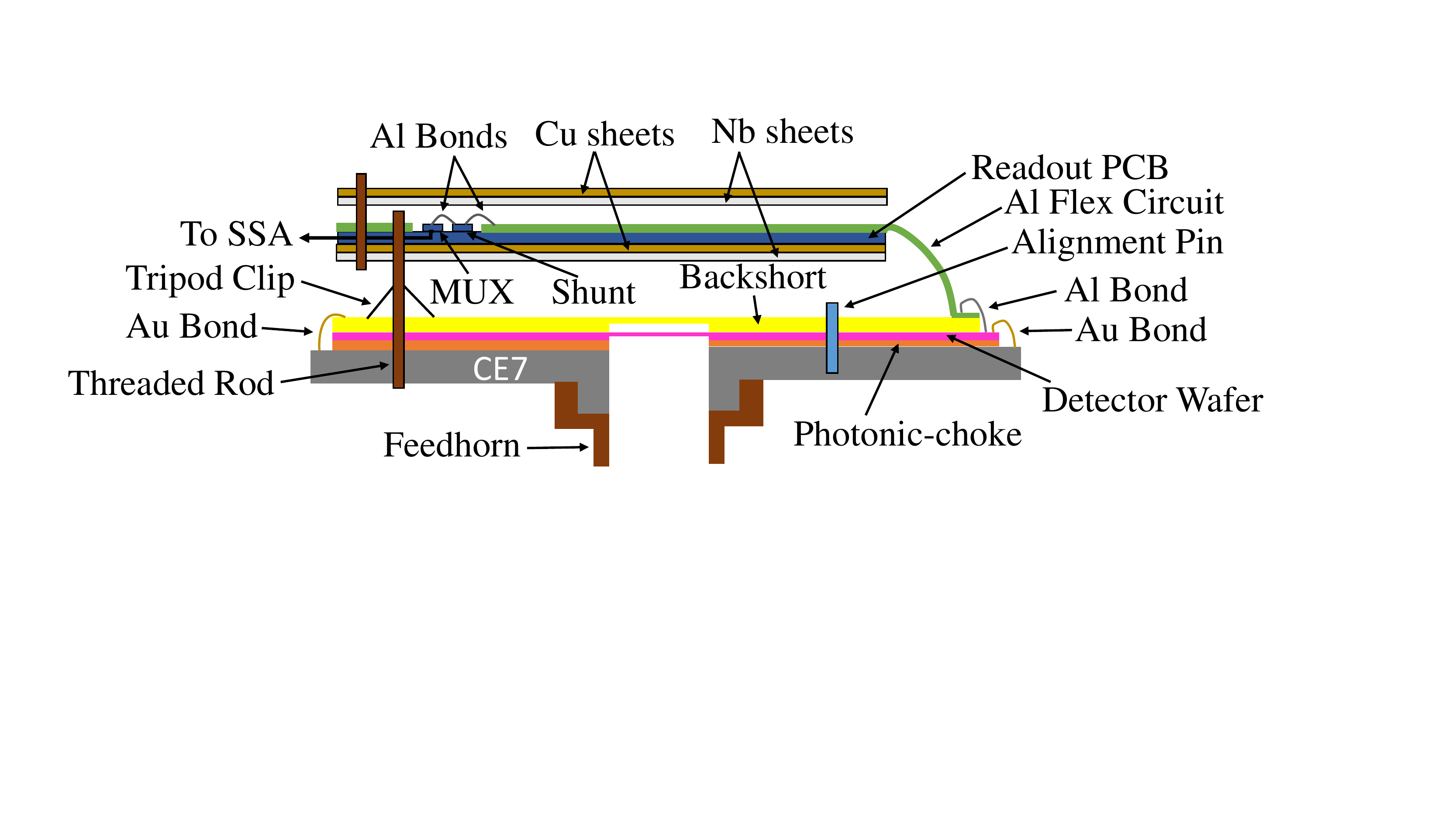}
   \end{tabular}
   \end{center}
   \caption[example] 
%>>>> use \label inside caption to get Fig. number with \ref{}
   { \label{fig:sketch} 
\textbf{Left:} 3D model showing cut sections of the W-band focal plane. Starting from CE7 baseplate and moving counter-clockwise, we show the detector wafer and readout circuit stack. First, the hybridized detector wafer is mounted on a Au-plated CE7 baseplate using three Be-Cu tripod spring clips and a side spring (not shown). The readout circuit with MUX and shunt chips on a PCB is then stacked on top of the detector wafer. The entire readout circuit assembly is sandwiched between two niobium sheets for magnetic shielding.  \textbf{Right:} Cross-section view (not to scale) of the detector chip and readout circuit stack. The sketch highlights how the base of a feedhorn mates with a cylindrical extrusion on the CE7 baseplate. The photonic-choke (orange), the detector wafer (pink), and the backshort assembly (yellow) are all hybridized during fabrication and mounted onto the baseplate as a single assembly. The top of the backshort assembly and the heat-sink pads on the detector wafer is gold bonded to the baseplate for heat sinking. The sketch also shows sets of aluminum bonds used to connect the detector bondpads to the readout circuit. Two copper sheets in the readout stack are connected to the backplate of each module for heat-sinking.}
   \end{sidewaysfigure}
   
Each module has a regular hexagonal baseplate with 50.8 mm sides. The baseplate is made from gold-plated CE7 (Controlled Expansion 7)\footnote{Sandvik Osprey, controlled-expansion CE7 alloy}  alloy, which is 70\% silicon and 30\% aluminum. CE7 is machinable and has lower differential contraction to silicon detector wafers than copper or aluminum \cite{aamir}. The coefficient of thermal expansion (CTE) mismatch between the baseplate and the wafer is a bigger concern for W-band as compared to Q-band where single-pixel detector chips were used \cite{karwan14_spie,john14}. CE7 is a poor thermal conductor below its superconducting transition temperature. Therefore, we gold plate it so that the detector wafer can be thermalised with the baseplate via gold bonds. These baseplates have 37 circular waveguide holes machined into them as shown in Figure \ref{fig:sketch}. The other side of the baseplate has matching cylindrical extrusions that mate with the slightly oversized cylindrical holes in the smooth-walled copper feedhorns \cite{zeng}. Due to the CTE mismatch between copper and CE7, the feedhorn base tightens against the cylindrical extrusion as the focal plane is cooled.

\begin{figure}[ht]
   \begin{center}
   \begin{tabular}{cc}
   \includegraphics[height=5.6cm]{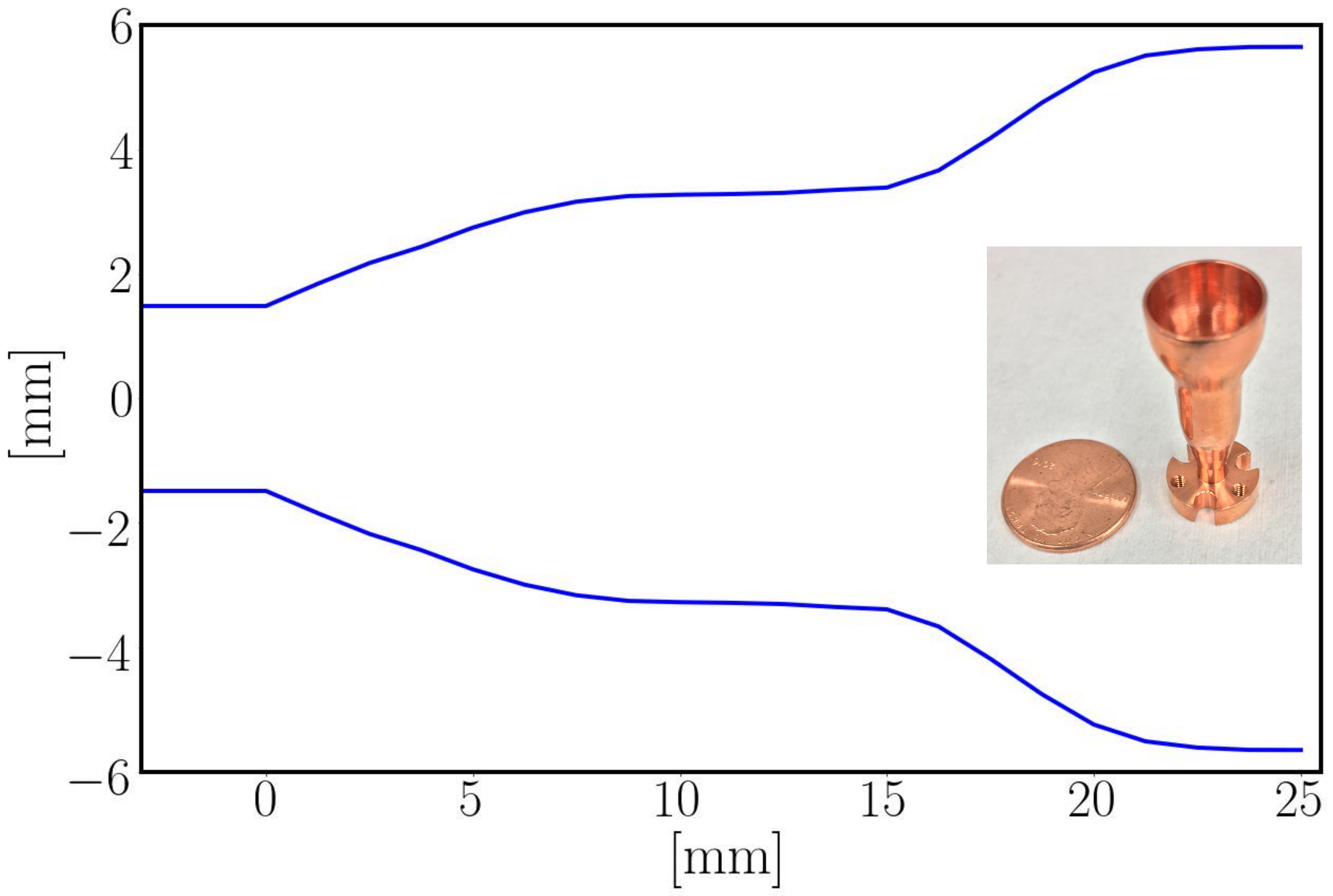}
   \includegraphics[height=5.6cm]{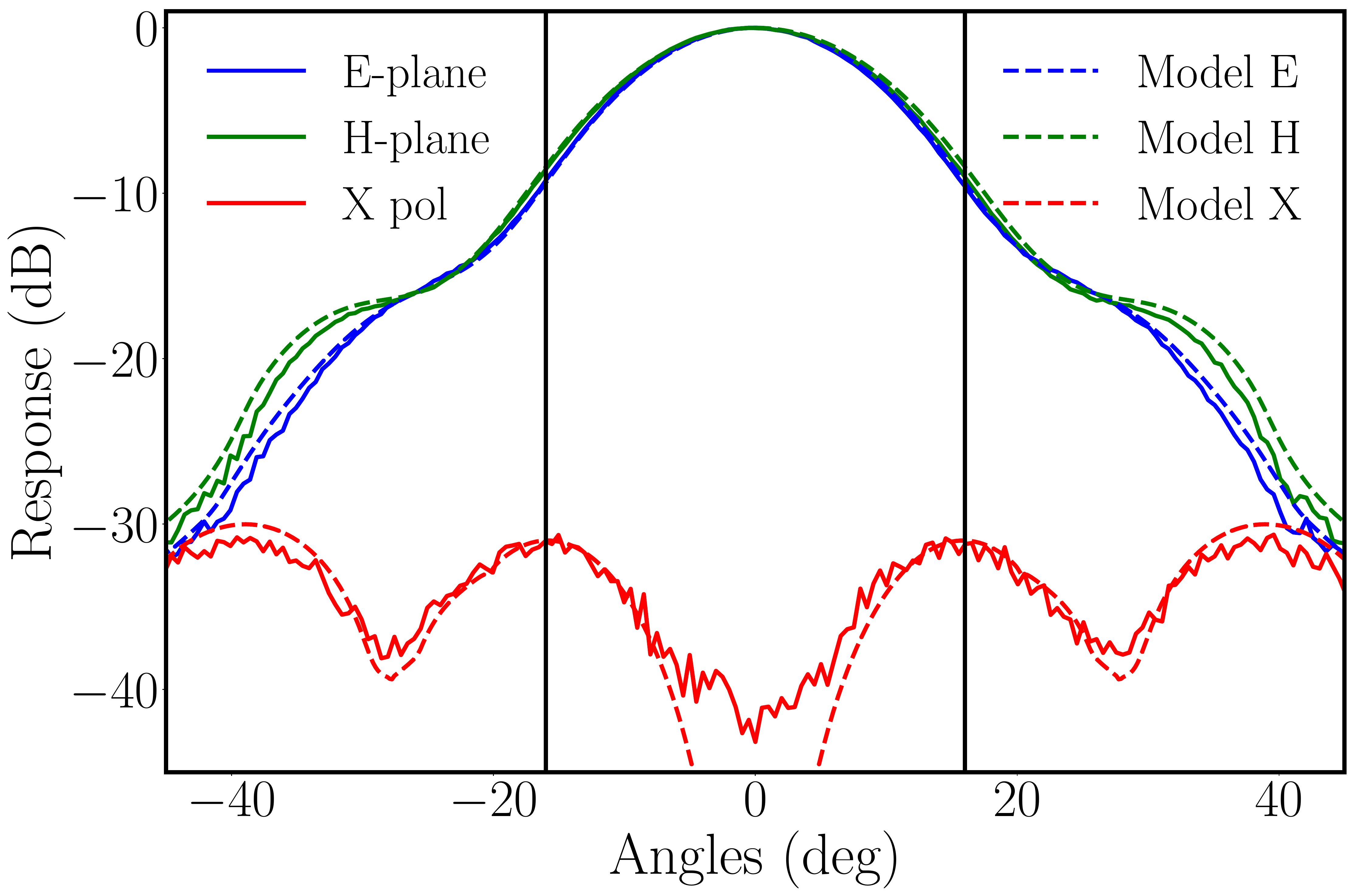}
   \end{tabular}
   \end{center}
   \caption[example] 
%>>>> use \label inside caption to get Fig. number with \ref{}
   { \label{fig:feed} \textbf{Left:} 20-point approximation of the W-band smooth-wall feedhorn profile. The inset shows a single feedhorn machined from oxygen-free high-conductivity copper. \textbf{Right:} W-band feedhorn co-polar E-plane and H-plane, and cross-polar, beam measurements averaged across the 77-108 GHz band, along with their models. The measurements were done in the Goddard Electromagnetic Anechoic Chamber (GEMAC) and show excellent agreement (within 2\%) of the models. The cross polarization across the bandpass is less than -30 dB, and the beam has a FWHM of $18.7^\circ$. The two vertical lines at $\pm 16^\circ$ show where the beams truncate on the receiver cold stop. The edge taper at 16$^\circ$ is $\approx$ -9 dB.}
   \end{figure}

CLASS W-band feedhorns are directly machined from oxygen-free high-conductivity (OFHC) copper. The smooth-wall profile of these feedhorns makes their manufacturing significantly less complex as compared to a corrugated profile. Figure \ref{fig:feed} shows the 20-point approximation of the W-band feedhorn profile that has an \linebreak 11.3 mm diameter aperture and a 2.97 mm throat that mates with the CE7 waveguide on the focal plane. We measured the beam pattern of the feedhorn in the Goddard Electromagnetic Anechoic Chamber (GEMAC). Figure \ref{fig:feed} shows the beam measurements averaged across the 77-108 GHz bandpass. The two vertical lines at $\pm 16^\circ$ show where the beams truncate on the receiver cold stop. The measurement shows that the cross polarization across the bandpass is less than -30 dB, and the beam has a FWHM of $18.7^\circ$. As seen in the figure, the measured beam pattern is in excellent agreement (within 2\%) with the modeled beam \cite{zeng}.           

The feedhorns and the baseplate waveguides couple light to the detectors mounted on the other side of the baseplate as shown in the cross-section sketch in Figure \ref{fig:sketch}.  The CLASS W-band detector chip consists of three separate wafers packaged together: a silicon photonic choke wafer, a monocrystalline silicon detector wafer, and a silicon backshort assembly. The photonic choke \cite{choke} wafer  acts as an interface between the baseplate and the detectors. This choke wafer consists of 1 $\mathrm{\mu}$m aluminum coated photonic choke pillars micro-machined by deep reactive ion etching (DRIE), and circular waveguides that align with the waveguide holes on the CE7 baseplate. The second wafer layer is the detector wafer with a 5 $\mathrm{\mu}$m thick float-zone single-crystal silicon layer serving as dielectric for microwave circuitry. Unlike the conventional use of silicon oxide or silicon nitride to simplify fabrication, CLASS uses this unique monocrystalline device layer for its excellent microwave, mechanical, and thermal properties \cite{karwan16}. The MoAu TES bilayer is deposited on this device layer. The final wafer layer is a backshort assembly \cite{backshort} that consists of a spacer wafer that is approximately a quarter-wave thick in-band and a cap wafer that forms a short for the orthomode transducer (OMT) termination. The CLASS-style detectors have been developed at NASA Goddard for space based application. For further description about the CLASS W-band wafer design and fabrication refer to Refs. \citenum {kevin}, \citenum{karwan16}, and \citenum{chuss}. 

This hybridized W-band detector wafer package is mounted as a single assembly onto the CE7 baseplate using three Be-Cu tripod spring clips as shown in Figure \ref{fig:sketch}. A custom designed step screw maintains $\sim$ 0.5 mm deflection of each clip, which puts sufficient force on the wafer to keep it stationary and ensure proper operation of the photonic choke-joints. Two alignment pins and a side mounted Be-Cu spring clip on the baseplate ensure proper alignment of the baseplate waveguides to the OMTs on the wafer. We vertically stack the readout circuit assembly on top of the step screws to enable the compact format of the module for its close packing in the focal plane. The readout assembly consists of eight time-domain multiplexer (MUX) \cite{nist, nist2} and shunt chips glued onto a printed circuit board (PCB) using rubber cement, and a flex circuit with Al traces stacked on top. All the MUX and shunt chips are fabricated by NIST. Each MUX chip contains 11 rows of Superconducting Quantum Interference Devices (SQUIDs), which are flux activated to select one detector to read out at a time. Each W-band module contains four pairs of MUX chips, with each pair strung together in order to multiplex over 22 rows. The shunt chips contain 250 $\mathrm{\mu} \Omega$ shunt resistors. CLASS W-band detector readout is described in detail in Section \ref{sec:det}. The detector bondpads are electrically connected to the MUX chips through the Al traces and shunt chips by putting Al bonds between each stage. The MUX chips are also bonded to the bondpads on the PCB. As shown in Figure \ref{fig:sketch}, the entire readout circuit assembly is sandwiched between a stack of 0.5 mm thick niobium (for magnetic shielding) and 0.1 mm thick copper sheets (for heat sinking) on both sides. The part of the PCB that protrudes out of the readout circuit assembly has twisted pairs of copper wires soldered onto it. These wires are connected to the PCB bondpads on one end, and 3 MDM connectors on the other: one MDM for the bias and feedback signals and two for multiplexing signals. The SQUID channels are connected to the SQUID series array (SSA) on a circuit board at the 4 K stage for signal amplification via NbTi superconducting cables. Using low-thermal-conductivity Manganin cables, the 4 K circuit board is then connected to the warm readout Multichannel Electronics (MCE) \cite{mce} provided by the University of British Columbia (UBC). All the wires are twisted pairs to reduce RF pick-up noise.

\begin{figure}[ht]
   \begin{flushleft}
   \begin{tabular}{c}
   \includegraphics[trim={5.2cm 2.7cm 4cm 2.6cm},clip,scale=0.69]{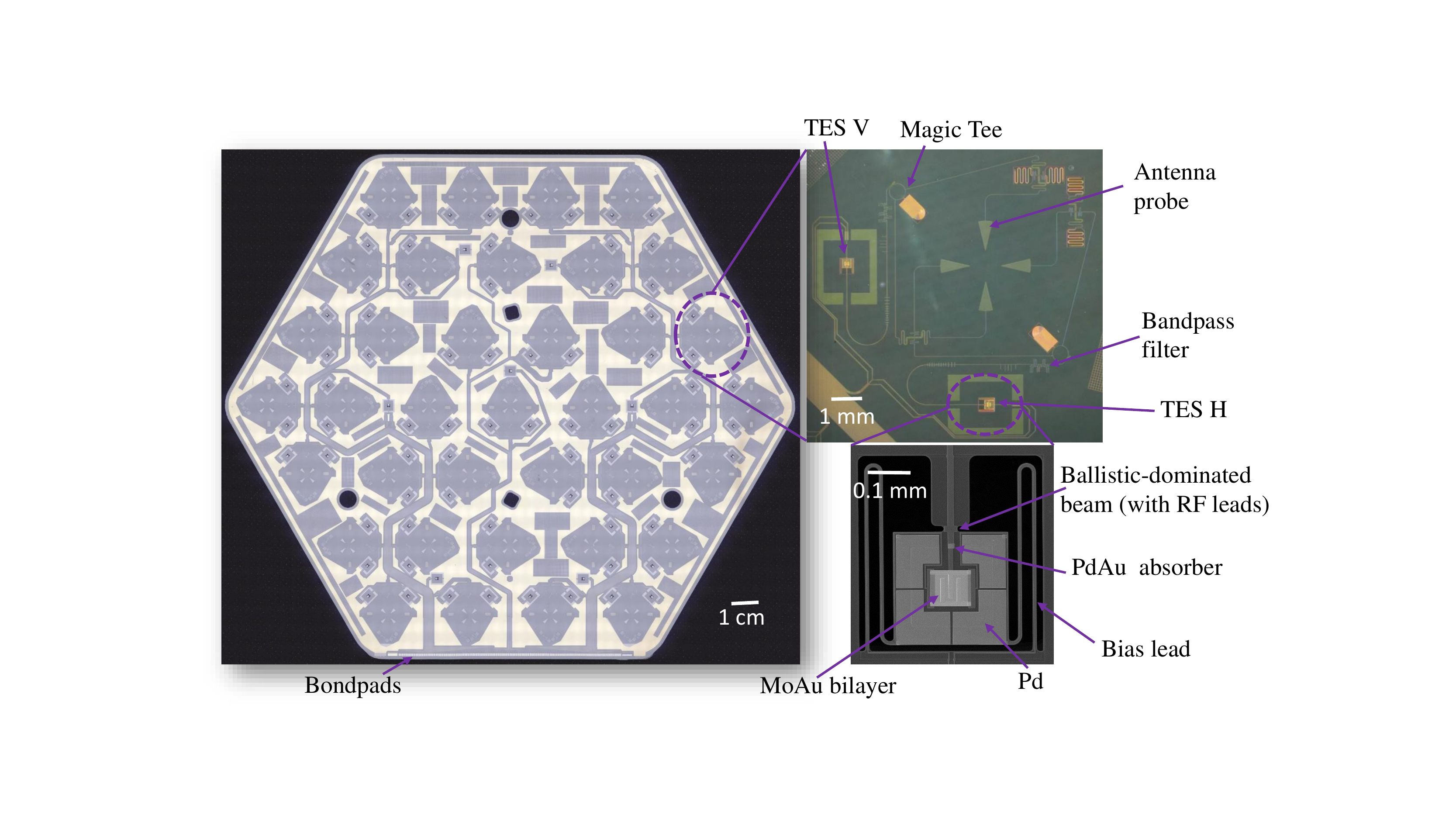}
   \end{tabular}
   \end{flushleft}
   \caption[example] 
%>>>> use \label inside caption to get Fig. number with \ref{}
   { \label{fig:Wafer} \textbf{Left:} CLASS W-band detector wafer showing the 37 dual-polarization-sensitive detectors. All the detectors are connected to the bondpads on the lower edge of the wafer. \textbf{Right:} Zoomed-in images of the detector circuit (top) and the TES island (bottom). For a detailed description about the W-band detector architecture refer to Ref. \citenum{karwan16}.}
   \end{figure}

\section{Detector Characterization}
\label{sec:det}
The CLASS W-band TES bolometers are made from a Mo-Au bilayer that is superconducting below $\sim$ 170 mK. The optical signal is  feedhorn-coupled to a planar membrane OMT that separates  orthogonal linear polarizations onto microstrip transmission lines terminated on the TES bolometers\cite{kevin,chuss} as shown in Figure \ref{fig:Wafer}. These TESs have normal resistance ($R_\mathrm{N}$) of $\sim$ 10 m$\Omega$ and are operated at $\sim$ 50\% $R_\mathrm{N}$. The TESs are stabilized through negative electrothermal feedback in a voltage biased circuit using a 250 $\mu\Omega$ shunt resister. The TES bilayer for CLASS detectors sits on an island that is thermally connected to the supporting structure through a set of silicon legs with weak thermal conductivity. The long and meandered legs, shown in the zoomed-in image of the TES island in Figure \ref{fig:Wafer}, support the TES bias leads. They have rough side walls, and therefore contribute very little to the thermal conductance. The thermal conductance of the island is precisely controlled by a short stubby beam with ballistic-dominated phonon transport \cite{karwan14}. For the W-band detectors, this stubby beam also carries the in-band microwave signal coming from the microstrips, which is terminated on a Pd-Au absorber thermally coupled to the TES as shown in Figure \ref{fig:Wafer}.

 \begin{figure}[b]
   \begin{center}
   \begin{tabular}{c} %% tabular useful for creating an array of images 
   \includegraphics[height=10cm]{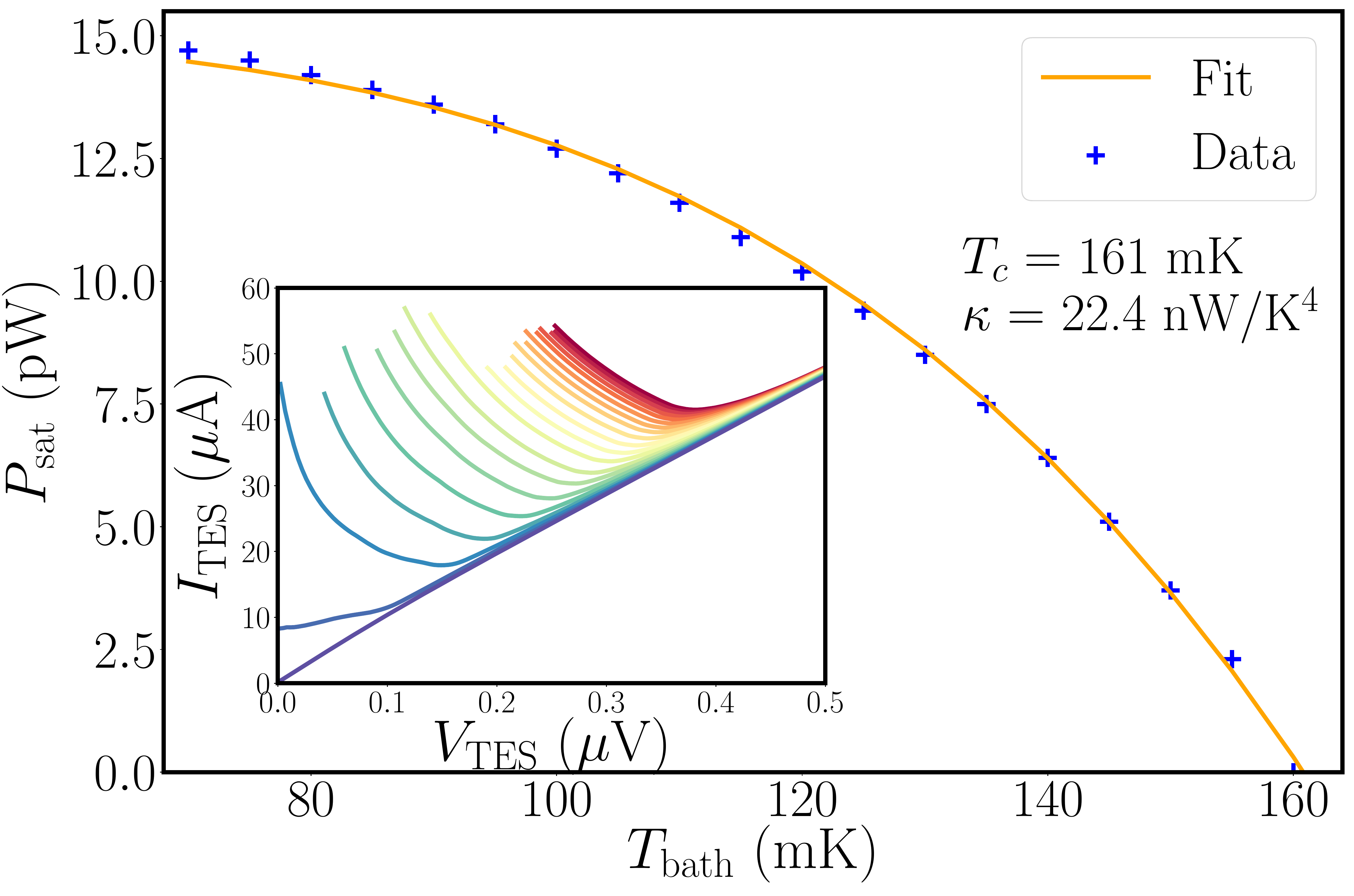}
   \end{tabular}
   \end{center}
   \caption[example] 
%>>>> use \label inside caption to get Fig. number with \ref{}
{ \label{fig:iv} The main plot shows the $P_\mathrm{sat}$ values obtained for one of the W-band detectors at multiple bath temperatures. The orange line shows the data fit to Equation \ref{eq:psat}. The fit for this particular detector gives $T_{\mathrm{c}}$ and $\kappa$ values of 161 mK and 22.4 nW/K$^4$ respectively. The inset shows the I-V curves used to calculate the $P_\mathrm{sat}$ values. The curves from red to blue correspond to bath temperatures from 70 mK to 165 mK with steps of 5 mK. Each curve terminates at the point where the TES becomes superconducting. }
    \end{figure}

\subsection{Dark Properties}
All the fabricated detector wafers go through extensive testing in order to verify that the expected parameters are close to target. We characterize the dark properties of the detectors by acquiring I-V curves for each detector at multiple bath temperatures. First, we ramp up the voltage bias to drive all the detectors normal, then we sweep the bias downwards over a wide range and record the current response of the detectors. As the bias voltage is lowered, the detector transitions from its normal to superconducting state. The I-V curves help us select the appropriate voltage bias to put the detector in transition while operating it in the field. The inset in Figure \ref{fig:iv} shows the I-V curves for one of the W-band detectors obtained at multiple bath temperatures. The curves are in increasing order of bath temperature from red to blue starting at 70 mK with temperature steps of 5 mK till 165 mK.

We acquire these I-V curves at multiple bath temperatures for all the detectors, and then calculate the TES saturation power ($P_\mathrm{sat}$) at each temperature through the product of the bias voltage ($V_\mathrm{TES}$) and response current ($I_\mathrm{TES}$) at the superconducting transition. $P_\mathrm{sat}$ is the amount of power required to raise the TES island temperature to the critical temperature ($T_{\mathrm{c}}$). Above its critical temperature, the TES is not sensitive to external signal, so $P_\mathrm{sat}$ is the maximum power a TES bolometer can measure. At a given bath temperature ($T_\mathrm{bath}$), $P_\mathrm{sat}$ can be calculated as:
\begin{equation}
\label{eq:psat}
P_{\mathrm{sat}} = \kappa \ (T_\mathrm{c}^n - T_{\mathrm{bath}}^n)\ ,
\end{equation}
where $\kappa$ is the parameter determined by the thermal conductance of the ballistic-dominated beam shown in Figure \ref{fig:Wafer}, and $n$ is the exponent governing the power flow between the TES island and the bath. In the ballistic phonon limit, which applies to CLASS detectors \cite{karwan14}, $n$ = 4.

\begin{figure}[ht]
   \begin{flushleft}
   \begin{tabular}{cc}
   \includegraphics[height=6.7cm]{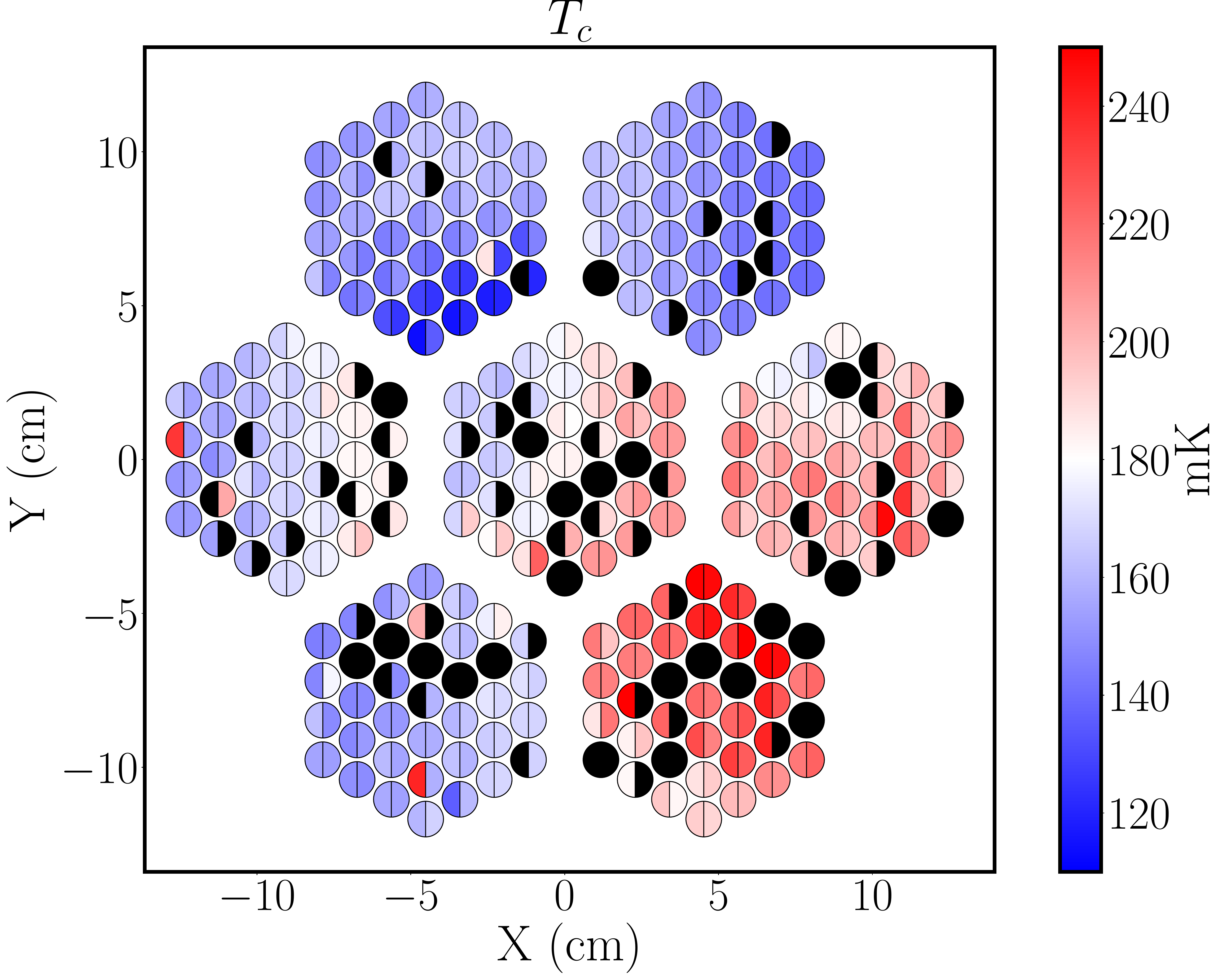}
   \includegraphics[height=6.7cm]{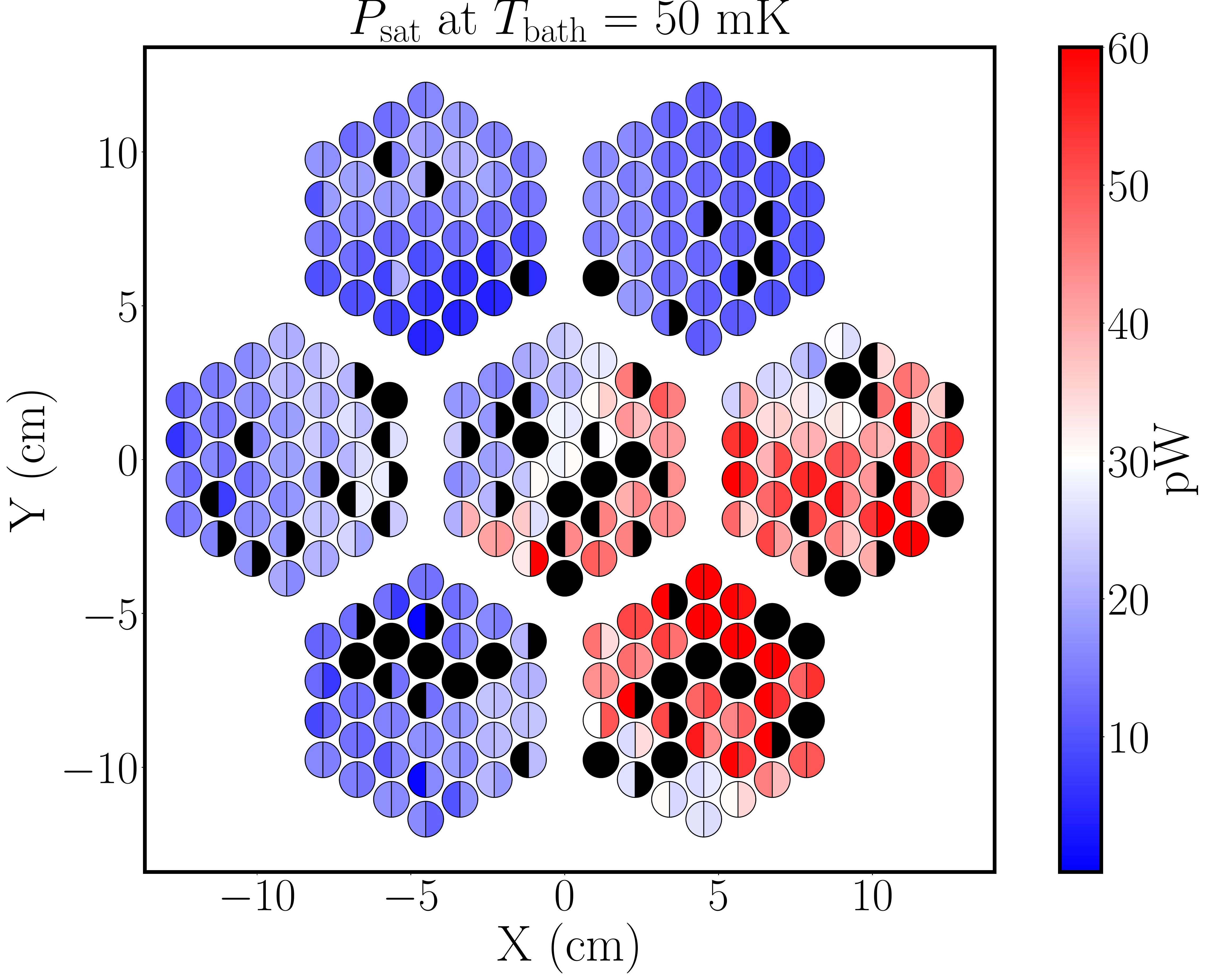}
   \end{tabular}
   \end{flushleft}
   \caption[example] 
%>>>> use \label inside caption to get Fig. number with \ref{}
   { \label{fig:wafer_plot} The distribution of $T_{\mathrm{c}}$ (left) and $P_\mathrm{sat}$ at $T_\mathrm{bath}$ = 50 mK (right) values for all the optically-sensitive detectors in the W-band focal plane. The X and Y axes represents the focal plane position compared to the detector at the center. The left and right sides of each circle show the H and V detectors respectively which are sensitive to separate orthogonal linear polarizations. The black spots show the detectors that did not show good enough I-V response to analyze. These plots show the current status of the first W-band detector array during deployment. In the future, we can improve TES uniformity across the focal plane by swapping modules in the current focal plane with the ones that are currently being tested in the lab.}
   \end{figure}

We fit $P_\mathrm{sat}$ vs $T_\mathrm{bath}$ to Equation \ref{eq:psat} to obtain $T_{\mathrm{c}}$ and $\kappa$ for each detector. Figure \ref{fig:iv} shows this fit for one of the W-band detectors obtained from multiple bath temperature I-V curves shown in the inset. For this particular detector, the curve fit gives $T_{\mathrm{c}}$ = 161 mK and $\kappa$ = 22.5 nW/K${^4}$. We perform this analysis for all the detectors in the focal plane. Figure \ref{fig:wafer_plot} shows the $T_{\mathrm{c}}$ and $P_\mathrm{sat}$ (at $T_\mathrm{bath}$ = 50 mK) values for all the optically-sensitive detectors in the first W-band focal plane. The left and right sides of each circle show the H and V detectors, respectively, which are sensitive to separate orthogonal linear polarizations. The black spots in the plot indicate the detectors whose I-V curves could not be obtained properly. In addition to these optically sensitive detectors, the W-band focal plane also contains several optically-isolated dark bolometers and dark SQUID channels. Dark bolometers can be used for probing light leaks and monitoring the stability of the bath temperature, whereas the dark SQUIDs can monitor readout noise and magnetic pick-up.

For an experiment like CLASS aiming to make a high-sensitivity measurement of the CMB polarization, it is critical that the $T_{\mathrm{c}}$ and $\kappa$ target values are chosen carefully. On one hand, lower $T_{\mathrm{c}}$ and $\kappa$  values lead to lower detector noise, which is described in detail in the following section. On the other hand, those values cannot be targeted too low; otherwise, they place the saturation power (described by Equation \ref{eq:psat}) below the optical loading. The target parameter values for the CLASS W-band detectors are listed in Table \ref{tab:det_param}. The table also includes the measured average parameter values for all the optically-sensitive CLASS detectors in the first W-band focal plane. The thermal conductance at the transition temperature ($G$), listed in the table, is related to $T_{\mathrm{c}}$ and $\kappa$ through:
\begin{equation}
\label{eq:G}
G = \dfrac{\mathrm{d}P_{\mathrm{sat}}}{\mathrm{d}T}\bigg|_{T_\mathrm{c}} = n\kappa T_\mathrm{c}^{n-1} \ .
\end{equation} 

Table \ref{tab:det_param} shows the total array yield of 82\%, which was calculated by counting the number of detectors that show good I-V response and fit Equation \ref{eq:psat} well. The remaining 18\% of the detectors are shown by the black spots in Figure \ref{fig:wafer_plot}. In addition to the parameters described so far, Table \ref{tab:det_param} also includes average $f_\mathrm{3dB}$ and $C$ values calculated using the effective detector time constant ($\tau_{\mathrm{eff}}$) measured from the response lag to a small square-wave excitation added to the detector bias voltage \cite{niemack}. $C$ is the TES heat capacity, and $f_\mathrm{3dB}$ is the frequency range where the power drops by less than half. $f_\mathrm{3dB}$ = 1/(2$\mathrm{\pi}\tau_{\mathrm{eff}})$, and  $\tau$ = $C/G$. We calculate $\tau$ by multiplying $\tau_{\mathrm{eff}}$ by the electro-thermal speed-up factor estimated from I-V curves. In the field, we use the VPM synchronous signal to get the actual optical detector time constant. This will be treated in detail in an upcoming publication. (J. W. Appel, et al., 2018, in preparation)  We calculated these values in lab to verify that they are close to our target values. \\

\begin{table}[ht]
\caption{First W-band Focal Plane average measured and target parameters} 
\label{tab:det_param}
\begin{center}       
\begin{tabular}{c|c|c} 
\rule[-1ex]{0pt}{3.5ex}  & \textbf{Measured} & \textbf{Target}  \\
\hline
\hline
\rule[-1ex]{0pt}{3.5ex}  $T_\mathrm{bath}$ & 35 mK &   \\
\hline
\rule[-1ex]{0pt}{3.5ex}  $T_{\mathrm{c}}$ & 175 mK & 150 mK  \\
\hline
\rule[-1ex]{0pt}{3.5ex}  $\kappa$ & 24.5 nW/K$^4$ & 25 nW/K$^4$ \\
\hline
\rule[-1ex]{0pt}{3.5ex}  $G$ @ $T_{\mathrm{c}}$ & 548 pW/K & 340 pW/K \\
\hline
\rule[-1ex]{0pt}{3.5ex} $P_\mathrm{sat}$ @ 50 mK  & 25 pW & 13 pW \\
\hline
\rule[-1ex]{0pt}{3.5ex}  $f_\mathrm{3dB}$ & 27 Hz & 30 Hz \\
\hline
\rule[-1ex]{0pt}{3.5ex}  $C$ & 4 pJ/K & 3 pJ/K  \\
\hline
\hline
\rule[-1ex]{0pt}{3.5ex}  $\textbf{Yield}$ & 82\% & \\
\end{tabular}
\end{center}
\end{table}

 \begin{figure} [ht]
   \begin{flushleft}
   \begin{tabular}{c} %% tabular useful for creating an array of images 
   \includegraphics[trim={3.5cm 0 4.5cm 1cm},clip,scale=0.4]{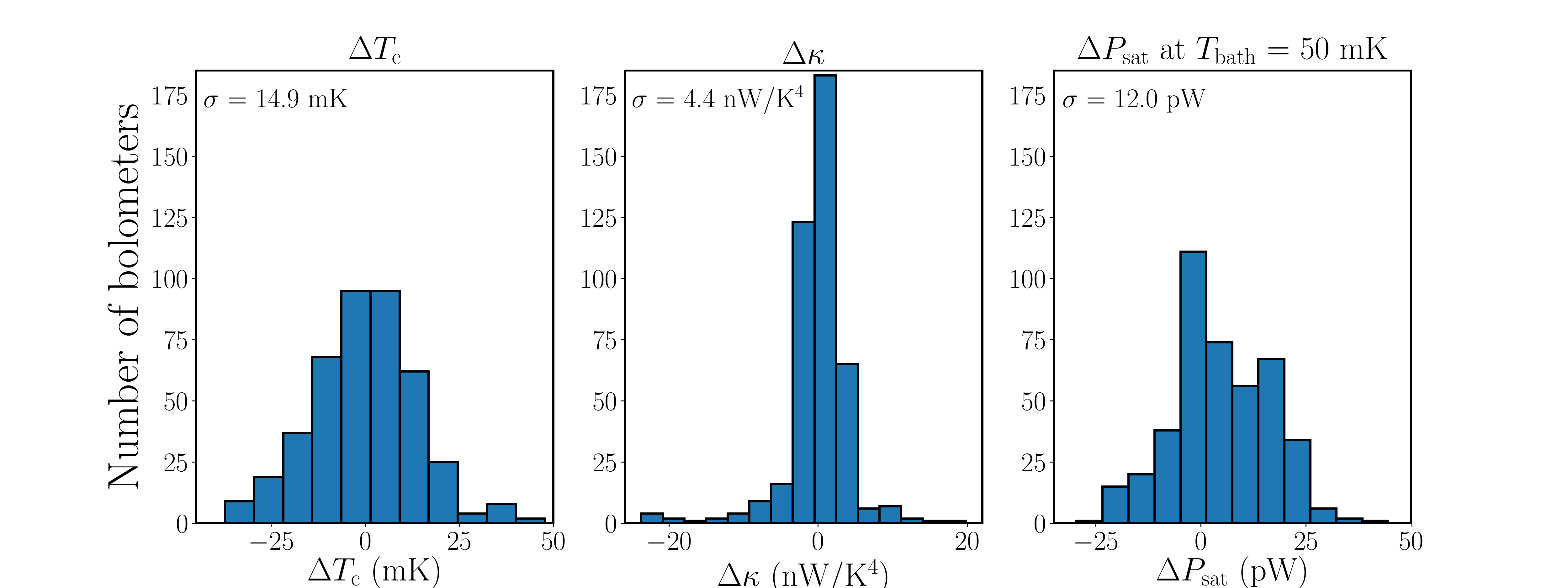}
   \end{tabular}
   \end{flushleft}
   \caption[example] 
%>>>> use \label inside caption to get Fig. number with \ref{}
   { \label{fig:hist} 
Histograms of $\Delta$$T_{\mathrm{c}}$, $\Delta \kappa$, and $\Delta$$P_\mathrm{sat}$ for 426 optically-sensitive detectors on the first W-band focal plane. These values were calculated by taking the difference of the individual detector parameter values with the average within their modules. Since each module has a separate detector bias line and all the detectors within a module share the same bias line, uniformity across $\Delta$ values reflects the optimal detector biasing condition. $P_\mathrm{sat}$ values were calculated at $T_\mathrm{bath}$ = 50 mK. \linebreak The $\sigma$ values on the upper left corner inside each box is the standard deviation for each distribution. }
   \end{figure}

From Figure \ref{fig:wafer_plot}, we can see that there are noticeable $T_{\mathrm{c}}$ and $P_{sat}$ gradients in  the detectors across the focal plane. However, although uniformity of individual detector parameters across the entire focal plane is desirable, it is critical that the detector parameters across individual wafers are uniform. This is because each W-band module has one detector bias line and all the detectors within a module have to be biased using one bias voltage. Figure \ref{fig:hist} shows the distribution of individual detector parameters as compared to the mean within their modules. The distribution includes values from 426 out of 518 optically-sensitive detectors in the focal plane (i.e. array yield of 82\%) with good I-V curves as described before. As shown in Figure \ref{fig:hist}, the variance in the thermal conductance parameter $\kappa$ across a wafer is very small ($\pm$ 18\%) which is a result of the ballistic thermal transport in each detector. The spread in $P_\mathrm{sat}$ ($\pm$ 47\%) is largely determined by the spread in $T_{\mathrm{c}}$ ($\pm$ 9\%). All the parameter distributions reported in this section show the current status of the first W-band detector array during deployment. In the future, we can improve TES uniformity across the focal plane by swapping modules in the current focal plane with the ones that are currently being tested in the lab. In addition, the development of AlMn-based TESs, which have shown to improve TES uniformity \cite{act}, is underway at NASA Goddard.

\subsection{Bandpass}
CLASS uses a combination of absorptive, reflective, and scattering filters inside the cryostat receiver to reject infrared radiation and reduce the out-of-band thermal loading on the focal plane \cite{jeff}. All these filters are tested and characterized in the lab for their out-of-band and in-band performance before being used in the field. The millimeter bandpass for CLASS detectors is defined through on-chip filtering \cite{chuss,kevin}. As shown in Figure \ref{fig:Wafer}, after the OMT couples the optical signal to the microstrip circuits, a series of thermal blocking and band-defining filters are used to both reject out-of-band radiation and precisely define the bandwidth. Therefore, it is crucial to test the bandpass of the detectors to check that the band-edges are on target and there is no optical power coupled at higher frequencies.

\begin{figure} [ht]
   \begin{center}
   \begin{tabular}{c} %% tabular useful for creating an array of images 
   \includegraphics[height=9cm]{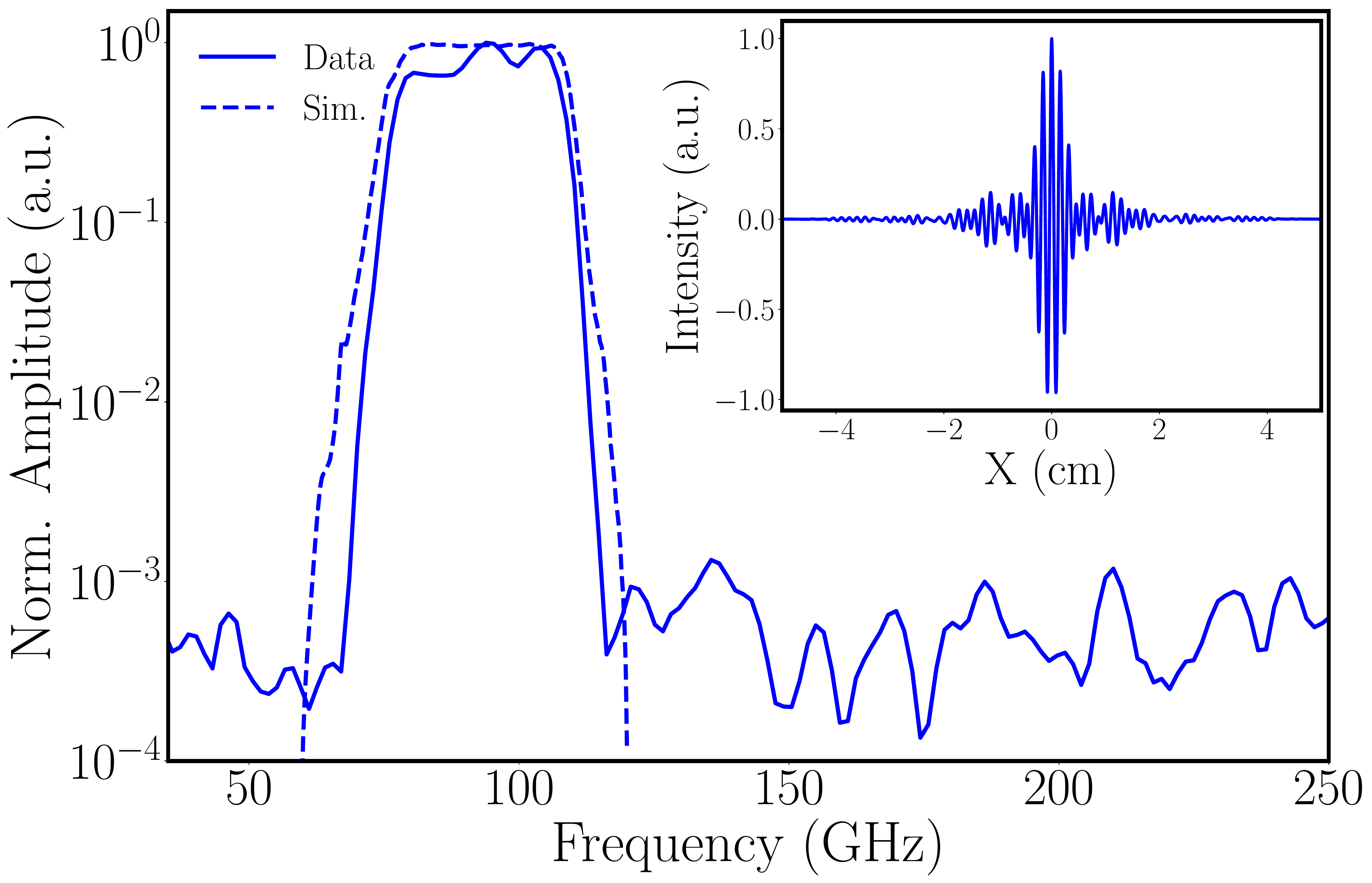}
   \end{tabular}
   \end{center}
   \caption[example] 
%>>>> use \label inside caption to get Fig. number with \ref{}
{ \label{fig:fts} The main plot shows the simulated and the measured bandpass of the W-band detectors measured using a Fourier Transform Spectrometer (FTS). The half-power points on the two band edges for this measurement are at 78 and 108 GHz, and the out-of-band response is less than -30 dB. This measured  bandpass is in good agreement with the simulation, and we see no evidence of optical power coupled at higher frequencies. The inset shows the apodized interferogram used to obtain the bandpass through a fast Fourier transform. The interferogram is a result of co-adding noise-weighted FTS signals from 21 detectors in one of the modules in the focal plane. The x-axis of the interferogram represents the position (centered at the white-light point) of the FTS movable mirror on a linear stage. The y-axis for both plots have been normalized to arbitrary units (a.u.).}
    \end{figure}

We use a Fourier Transform Spectrometer (FTS) to measure the bandpass of CLASS W-band detectors. To not saturate the detectors with the FTS thermal source, the receiver was covered with a metal plate with a \linebreak 5 cm diameter hole in the center. In addition, anti-reflection coated Teflon and Nylon filters were placed at the 60 K and 4 K stages of the cryostat respectively. As the FTS movable mirror scans back and forth from 0 to 150 mm at 0.5 mm/s on a linear stage, the detectors measure the output radiation intensity. We chopped the FTS wide-band thermal source at 20 Hz to modulate the FTS signal. The inset in Figure \ref{fig:fts} shows an apodized interferogram obtained from detectors in one of the modules in the W-band focal plane. This interferogram is the result of co-adding noise-weighted FTS signals from 21 detectors in the module. The real component of the Fourier transform of the interferogram yields the bandpass shown in the main plot. The half-power points on the two band edges for this measurement are at 78 and 108 GHz, and the out-of-band response is less than -30 dB. This measured W-band detector bandpass is in good agreement with the simulation as seen in Figure \ref{fig:fts}, and we see no evidence of optical power coupled at higher frequencies.         

\subsection{Noise Performance}
The total array noise-equivalent power (NEP) of a focal plane with N detectors scales as 1/$\sqrt[]{\mathrm{N}}$ with the NEP of individual detectors. CLASS detectors are designed to be photon noise limited; therefore, we can only achieve higher sensitivity by increasing the number of detectors. Phonon noise from the thermal link, Johnson noise from the detector circuit resistance, and SQUID amplifier noise all contribute to the so-called ``dark NEP'', which is measured with the cryostat closed so as to minimize noise from background radiation. The SQUID NEP for CLASS TES bolometers is expected to be $\sim$ 3 aW$\sqrt[]{\mathrm{s}}$ \cite{john14}. When adding this NEP in quadrature to NEP from other noise sources, it contributes less than a few percent of the total noise. The Johnson noise is suppressed at low frequencies due to electro-thermal feedback \cite{Irwin}. The phonon noise depends strongly on $T_{\mathrm{c}}$ as follows:   
\begin{equation}
\label{eq:ph_noise}
\mathrm{NEP_{phonon}} = \sqrt[]{2k_\mathrm{B}T_\mathrm{c}^2GF_\mathrm{link}}\ ,
\end{equation}
where NEP has units of W$\sqrt[]{\mathrm{s}}$, $k_\mathrm{B}$ is the Boltzmann constant, and $F_\mathrm{link}$ depends on the energy transport mechanism between the TES island and the bath. For ballistic phonon transport, its value is given by (1+ ($T_{\mathrm{bath}}$/$T_{\mathrm{c}})^{n+1}$)/2 \cite{boyle}. Given the low $T_{\mathrm{bath}}$ as compared to $T_{\mathrm{c}}$ for the CLASS detectors, $F_\mathrm{link}$ = 1/2. Due to the strong dependence of detector NEP on $T_{\mathrm{c}}$ and $\kappa$ (from $G$), the CLASS TES bilayer and the geometry of the TES legs were carefully designed to optimize detector dark NEP \cite{john14}.

 \begin{figure} [ht]
   \begin{center}
   \begin{tabular}{c} %% tabular useful for creating an array of images 
   \includegraphics[height=9cm]{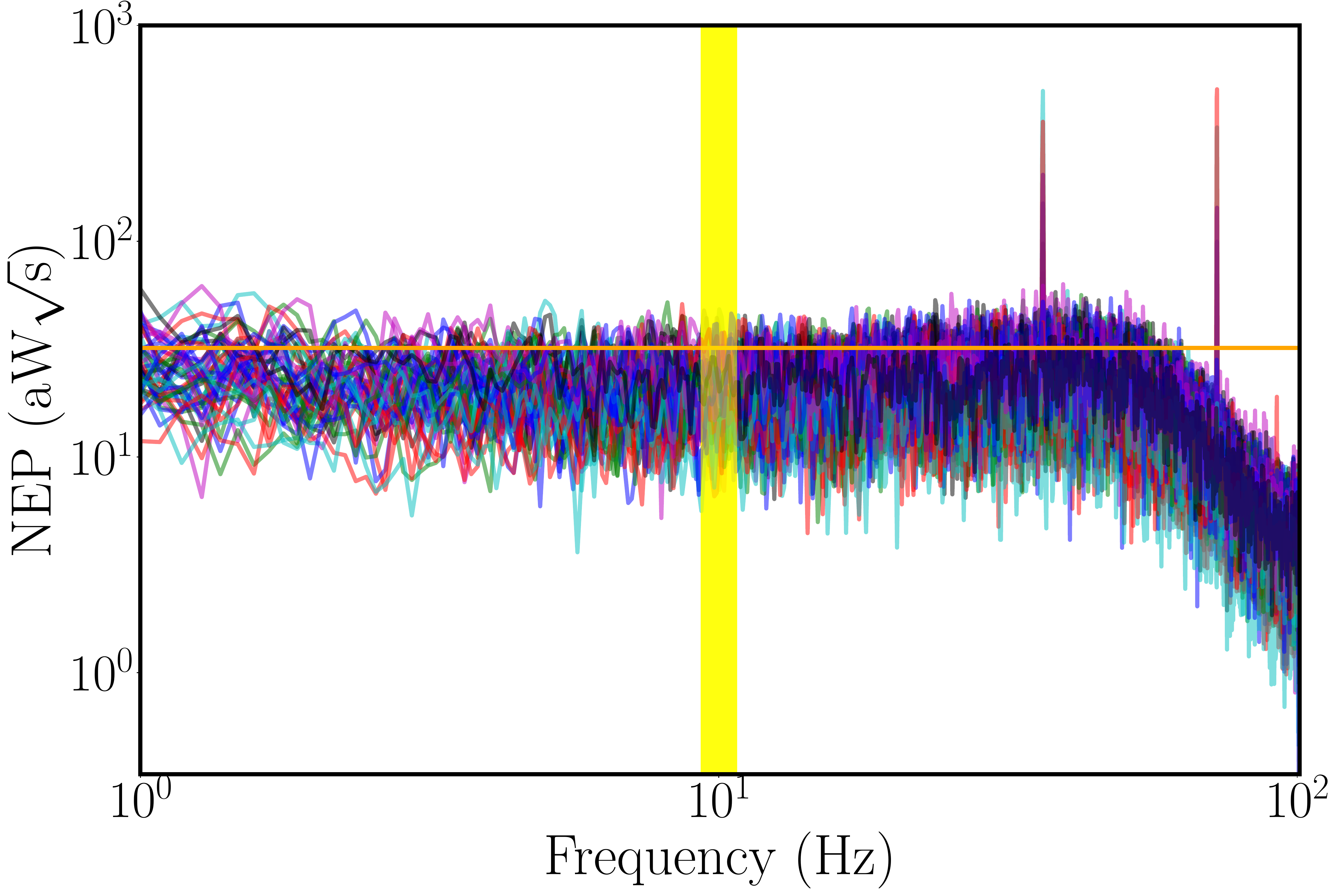}
   \end{tabular}
   \end{center}
   \caption[example] 
%>>>> use \label inside caption to get Fig. number with \ref{}
   { \label{fig:nep} Noise spectra of 48 science-grade detectors in one of the modules in the W-band focal plane. CLASS signal band is shown by the vertical yellow patch centered at the VPM modulation frequency of 10 Hz. The horizontal orange line indicates an estimated photon NEP in the field of 32 aW$\sqrt[]{\mathrm{s}}$. The total NEP for CLASS detectors is dominated by photon noise.}
    \end{figure}

To characterize the dark NEP of the detectors, we took noise spectra of individual modules capping off all the cold stages of the cryostat receiver with metal plates. Figure \ref{fig:nep} shows the noise spectra of one of the modules in the W-band focal plane. The plot includes spectra from 48 science-grade detectors in the module. The vertical yellow band indicates the VPM modulation frequency of 10 Hz. This modulation puts our signal band away from 1/$f$ noise at low frequency, which comes from a combination of instrumental and atmospheric drifts. The roll-off observed at $\sim$ 60 Hz is due to the MCE digital Butterworth filter applied to the readout to suppress noise aliasing from higher frequencies. The horizontal orange line indicates an estimated photon NEP in the field of 32 aW$\sqrt[]{\mathrm{s}}$ \cite{tom14}. The photon noise comes from a combination of the CMB and emission from the telescope and atmosphere. The total detector NEP, which is a sum of the detector dark NEP and the photon NEP added in quadrature, for CLASS detectors is dominated by the photon NEP as shown in Figure \ref{fig:nep}. At around 10 Hz, the mean dark NEP of all the 48 detectors shown in the spectra is 21 aW$\sqrt[]{\mathrm{s}}$. Therefore, we expect the total NEP of the W-band detectors to be 38 aW$\sqrt[]{\mathrm{s}}$.     

\subsection{Sensitivity Projections}
Based on the detector parameters presented in this paper, we can estimate the sensitivity for the first W-band focal plane. For this calculation, we use the mean total NEP per detector of 38 aW$\sqrt[]{\mathrm{s}}$. 426 detectors on the focal plane showed good I-V response.  However, due to the spread in $T_{\mathrm{c}}$ and $P_\mathrm{sat}$ and having only one detector bias line per module, we might not be able to bias all the working detectors in the field simultaneously. We take a conservative estimate that we can only bias 90\% of those detectors, i.e. 383 detectors (array efficiency of $\sim$ 74\%). This puts the total array NEP for the first CLASS W-band focal plane at 2.1 aW$\sqrt[]{\mathrm{s}}$.

For the 77-108 GHz CLASS bandpass, the conversion factor from sky power to CMB temperature (d$P$/d$T_\mathrm{cmb}$) is 0.34 pWK$^{-1}$. We are currently working on measuring the detector efficiency using a waveguide thermal source \cite{karwan13}. For this sensitivity estimate, we use a preliminary detector efficiency estimate of 70\% and estimate the remaining telescope optics to have a nominal efficiency of $\sim$ 62\%. Refer to Ref. \citenum{tom14} for more details on the telescope optical efficiency estimate. Now, the total array noise-equivalent CMB temperature (NET$_\mathrm{cmb}$) can be calculated as:
\begin{equation}
\label{eq:NET}
\mathrm{NET_{cmb} = \dfrac{NEP_{array}}{\epsilon_{total}}} \dfrac{\mathrm{d}T_{cmb}}{\mathrm{d}P} = 14\  \mu \mathrm{K}\ \sqrt[]{\mathrm{s}}  \ ,
\end{equation}
where $\epsilon_\mathrm{total}$ = detector efficiency $\times$ telescope optics efficiency. Assuming VPM modulation efficiency of 70\% to measure the Stokes parameter Q, this array NET translates to noise-equivalent Q (NEQ) of  20\  $\mu$K$\ \sqrt[]{\mathrm{s}}$.  

\section{Conclusion}
\label{sec:conclusion}
The CLASS Q-band receiver has been operational in the field since May 2016, and the first W-band receiver has recently been deployed. All CLASS detectors are feedhorn-coupled transition-edge bolometers that couple to light using planar ortho-mode transducer antennas and smooth-walled feedhorns. The W-band focal plane detector array consists of seven individual modules with 37 feedhorns placed on a CE7 baseplate with cylindrical waveguide holes machined in them. Each module contains 37 dual-polarization sensitive detectors fabricated on a monocrystalline silicon used as a dielectric substrate. Before being fielded, all the modules on the first W-band focal plane were extensively tested and characterized for their parameter uniformity and noise performance. The measured $T_{\mathrm{c}}$, $\kappa$, and $P_\mathrm{sat}$ values are within margins targeted for the W-band receiver. The in-lab FTS measurement shows that the detector bandpass is in good agreement with the simulation, and we see no evidence of optical power coupling at higher frequencies. We measure the mean dark NEP for science-grade detectors in a module to be 21 aW$\sqrt[]{\mathrm{s}}$, and estimate the total array NEP of the focal plane to be 2.1 aW$\sqrt[]{\mathrm{s}}$. Using the detector parameters presented in this paper, we estimate the total array NET to be 14 $\mu$K$\ \sqrt[]{\mathrm{s}}$, which translates to total array NEQ of 20 $\mu$K$\ \sqrt[]{\mathrm{s}}$, assuming 70\% VPM modulation efficiency. The second W-band focal plane has the same design, and the detector wafers for it are being fabricated and tested. The CLASS W-band detector arrays are optimized for CMB observations near the minimum of polarized Galactic emission.

\acknowledgments % equivalent to \section*{ACKNOWLEDGMENTS}       

Among the four lead authors of this paper, S. Dahal is a graduate student at JHU who completed the W-band focal plane assembly and detector characterization and installed and verified the detector array at the CLASS telescope site in the Atacama Desert. A. Ali is a postdoctoral researcher at UC Berkeley whose JHU-based PhD dissertation encompassed the bulk of the W-band focal plane development, including assembly, testing of detectors, characterization of CE7, and magnetic shielding design \cite{aamir}. J. W. Appel is an Associate Research Scientist who has overseen all CLASS detector testing, assembly and in-field characterization.   
T. Essinger-Hileman is a Research Astrophysicist at NASA Goddard who designed the focal plane modules, including the use of CE7 as a silicon interface material. 

We acknowledge the National Science Foundation Division of Astronomical Sciences for their support of CLASS under Grant Numbers 0959349, 1429236, 1636634, and 1654494. CLASS uses detector technology developed under several previous and ongoing NASA grants. Detector development work at JHU was funded by NASA grant number NNX14AB76A.  We are also grateful to NASA for their support of civil servants engaged in state-of-the-art detector technologies. K. Harrington is supported by the NASA Space Technology Research Fellowship grant number NXX14AM49H. T. Essinger-Hileman was supported by an NSF Astronomy and Astrophysics Postdoctoral Fellowship. We further acknowledge the very generous support of Jim and Heather Murren (JHU A\&S '88), Matthew Polk (JHU A\&S Physics BS '71), David Nicholson, and Michael Bloomberg (JHU Engineering  '64). CLASS is located in the Parque Astron\'{o}mico de Atacama in northern Chile under the auspices of the Comisi\'{o}n Nacional de Investigaci\'{o}n Cient\'{i}fica y Tecnol\'{o}gica de Chile (CONICYT).

% References
\bibliography{report} % bibliography data in report.bib

\begin{thebibliography}{10}

\bibitem{fixsen}
{Fixsen}, D.~J., ``{The Temperature of the Cosmic Microwave Background},'' {\em
  \apj}~{\bf 707},  916--920 (Dec. 2009).

\bibitem{wmap}
{Bennett}, C.~L., {Larson}, D., {Weiland}, J.~L., {Jarosik}, N., {Hinshaw}, G.,
  {Odegard}, N., {Smith}, K.~M., {Hill}, R.~S., {Gold}, B., {Halpern}, M.,
  {Komatsu}, E., {Nolta}, M.~R., {Page}, L., {Spergel}, D.~N., {Wollack}, E.,
  {Dunkley}, J., {Kogut}, A., {Limon}, M., {Meyer}, S.~S., {Tucker}, G.~S., and
  {Wright}, E.~L., ``{Nine-year Wilkinson Microwave Anisotropy Probe (WMAP)
  Observations: Final Maps and Results},'' {\em \apjs}~{\bf 208},  20 (Oct.
  2013).

\bibitem{planck}
{Planck Collaboration}, {Ade}, P.~A.~R., {Aghanim}, N., {Arnaud}, M.,
  {Ashdown}, M., {Aumont}, J., {Baccigalupi}, C., {Banday}, A.~J., {Barreiro},
  R.~B., {Bartlett}, J.~G., and et~al., ``{Planck 2015 results. XIII.
  Cosmological parameters},'' {\em \aap}~{\bf 594},  A13 (Sept. 2016).

\bibitem{guth}
{Guth}, A.~H., ``{Inflationary universe: A possible solution to the horizon and
  flatness problems},'' {\em \prd}~{\bf 23},  347--356 (Jan. 1981).

\bibitem{linde}
{Linde}, A.~D., ``{A new inflationary universe scenario: A possible solution of
  the horizon, flatness, homogeneity, isotropy and primordial monopole
  problems},'' {\em Physics Letters B}~{\bf 108},  389--393 (Feb. 1982).

\bibitem{marc}
{Kamionkowski}, M., {Kosowsky}, A., and {Stebbins}, A., ``{Statistics of cosmic
  microwave background polarization},'' {\em \prd}~{\bf 55},  7368--7388 (June
  1997).

\bibitem{zaldarriaga}
{Zaldarriaga}, M. and {Seljak}, U., ``{All-sky analysis of polarization in the
  microwave background},'' {\em \prd}~{\bf 55},  1830--1840 (Feb. 1997).

\bibitem{watts15}
{Watts}, D.~J., {Larson}, D., {Marriage}, T.~A., {Abitbol}, M.~H., {Appel},
  J.~W., {Bennett}, C.~L., {Chuss}, D.~T., {Eimer}, J.~R., {Essinger-Hileman},
  T., {Miller}, N.~J., {Rostem}, K., and {Wollack}, E.~J., ``{Measuring the
  Largest Angular Scale CMB B-mode Polarization with Galactic Foregrounds on a
  Cut Sky},'' {\em \apj}~{\bf 814},  103 (Dec. 2015).

\bibitem{tom14}
{Essinger-Hileman}, T., {Ali}, A., {Amiri}, M., {Appel}, J.~W., {Araujo}, D.,
  {Bennett}, C.~L., {Boone}, F., {Chan}, M., {Cho}, H.-M., {Chuss}, D.~T.,
  {Colazo}, F., {Crowe}, E., {Denis}, K., {D{\"u}nner}, R., {Eimer}, J.,
  {Gothe}, D., {Halpern}, M., {Harrington}, K., {Hilton}, G.~C., {Hinshaw},
  G.~F., {Huang}, C., {Irwin}, K., {Jones}, G., {Karakla}, J., {Kogut}, A.~J.,
  {Larson}, D., {Limon}, M., {Lowry}, L., {Marriage}, T., {Mehrle}, N.,
  {Miller}, A.~D., {Miller}, N., {Moseley}, S.~H., {Novak}, G., {Reintsema},
  C., {Rostem}, K., {Stevenson}, T., {Towner}, D., {U-Yen}, K., {Wagner}, E.,
  {Watts}, D., {Wollack}, E.~J., {Xu}, Z., and {Zeng}, L., ``{CLASS: the
  cosmology large angular scale surveyor},'' in [{\em Proceedings of the SPIE,
  Volume 9153, id. 91531I 23 pp. (2014).}{\nolinebreak\hspace{0.1em}]},   {\bf
  9153} (July 2014).

\bibitem{katie}
{Harrington}, K., {Marriage}, T., {Ali}, A., {Appel}, J.~W., {Bennett}, C.~L.,
  {Boone}, F., {Brewer}, M., {Chan}, M., {Chuss}, D.~T., {Colazo}, F., {Dahal},
  S., {Denis}, K., {D{\"u}nner}, R., {Eimer}, J., {Essinger-Hileman}, T.,
  {Fluxa}, P., {Halpern}, M., {Hilton}, G., {Hinshaw}, G.~F., {Hubmayr}, J.,
  {Iuliano}, J., {Karakla}, J., {McMahon}, J., {Miller}, N.~T., {Moseley},
  S.~H., {Palma}, G., {Parker}, L., {Petroff}, M., {Pradenas}, B., {Rostem},
  K., {Sagliocca}, M., {Valle}, D., {Watts}, D., {Wollack}, E., {Xu}, Z., and
  {Zeng}, L., ``{The Cosmology Large Angular Scale Surveyor},'' in [{\em
  Millimeter, Submillimeter, and Far-Infrared Detectors and Instrumentation for
  Astronomy VIII}{\nolinebreak\hspace{0.1em}]},  {\em \procspie} {\bf 9914},
  99141K (July 2016).

\bibitem{watts18}
{Watts}, D.~J., {Wang}, B., {Ali}, A., {Appel}, J.~W., {Bennett}, C.~L.,
  {Chuss}, D.~T., {Dahal}, S., {Eimer}, J.~R., {Essinger-Hileman}, T.,
  {Harrington}, K., {Hinshaw}, G., {Iuliano}, J., {Marriage}, T.~A., {Miller},
  N.~J., {Padilla}, I.~L., {Petroff}, M., {Rostem}, K., {Wollack}, E.~J., and
  {Xu}, Z., ``{A Projected Estimate of the Reionization Optical Depth Using the
  CLASS Experiment's Sample-Variance Limited E-Mode Measurement},'' {\em ArXiv
  e-prints}  (Jan. 2018).

\bibitem{ricardo}
Bustos, R., Rubio, M., Otárola, A., and Nagar, N., ``Parque astronómico de
  atacama: An ideal site for millimeter, submillimeter, and mid-infrared
  astronomy,'' {\em Publications of the Astronomical Society of the
  Pacific}~{\bf 126}(946),  1126 (2014).

\bibitem{nathan}
{Miller}, N.~J., {Chuss}, D.~T., {Marriage}, T.~A., {Wollack}, E.~J., {Appel},
  J.~W., {Bennett}, C.~L., {Eimer}, J., {Essinger-Hileman}, T., {Fixsen},
  D.~J., {Harrington}, K., {Moseley}, S.~H., {Rostem}, K., {Switzer}, E.~R.,
  and {Watts}, D.~J., ``{Recovery of Large Angular Scale CMB Polarization for
  Instruments Employing Variable-delay Polarization Modulators},'' {\em
  \apj}~{\bf 818},  151 (Feb. 2016).

\bibitem{katie18}
{Harrington}, K., {Eimer}, J., {Chuss}, D.~T., {Petroff}, M., {Cleary}, J.,
  {DeGeorge}, M., {Ali}, A., {Appel}, J.~W., {Bennett}, C.~L., {Brewer}, M.,
  {Bustos}, R., {Chan}, M., {Couto}, J., {Dahal}, S., {Denis}, K.,
  {D{\"u}nner}, R., {Essinger-Hileman}, T., {Fluxa}, P., {Halpern}, M.,
  {Hilton}, G., {Hinshaw}, G.~F., {Hubmayr}, J., {Iuliano}, J., {Karakla}, J.,
  {Marriage}, T., {McMahon}, J., {Miller}, N.~T., {Nu\~{n}ez}, C., {Padilla},
  I.~L., {Palma}, G., {Parker}, L., {Pradenas}, B., {Reeves}, R., {Reintsema},
  C., {Rostem}, K., {Valle}, D., {Van Engelhoven}, T., {Wang}, B., {Wang}, Q.,
  {Watts}, D., {Weiland}, J., {Wollack}, E., {Xu}, Z., {Yan}, Z., and {Zeng},
  L., ``{Variable-delay Polarization Modulators for the for the CLASS
  Telescopes},'' in [{\em Millimeter, Submillimeter, and Far-Infrared Detectors
  and Instrumentation for Astronomy IX}{\nolinebreak\hspace{0.1em}]},  {\em
  \procspie} {\bf 10708},  10708--92 (July 2018).

\bibitem{joseph}
{Eimer}, J.~R., {Bennett}, C.~L., {Chuss}, D.~T., {Marriage}, T., {Wollack},
  E.~J., and {Zeng}, L., ``{The cosmology large angular scale surveyor (CLASS):
  40 GHz optical design},'' in [{\em Millimeter, Submillimeter, and
  Far-Infrared Detectors and Instrumentation for Astronomy
  VI}{\nolinebreak\hspace{0.1em}]},  {\em \procspie} {\bf 8452},  845220 (Sept.
  2012).

\bibitem{john14}
{Appel}, J.~W., {Ali}, A., {Amiri}, M., {Araujo}, D., {Bennet}, C.~L., {Boone},
  F., {Chan}, M., {Cho}, H.-M., {Chuss}, D.~T., {Colazo}, F., {Crowe}, E.,
  {Denis}, K., {D{\"u}nner}, R., {Eimer}, J., {Essinger-Hileman}, T., {Gothe},
  D., {Halpern}, M., {Harrington}, K., {Hilton}, G., {Hinshaw}, G.~F., {Huang},
  C., {Irwin}, K., {Jones}, G., {Karakula}, J., {Kogut}, A.~J., {Larson}, D.,
  {Limon}, M., {Lowry}, L., {Marriage}, T., {Mehrle}, N., {Miller}, A.~D.,
  {Miller}, N., {Moseley}, S.~H., {Novak}, G., {Reintsema}, C., {Rostem}, K.,
  {Stevenson}, T., {Towner}, D., {U-Yen}, K., {Wagner}, E., {Watts}, D.,
  {Wollack}, E., {Xu}, Z., and {Zeng}, L., ``{The cosmology large angular scale
  surveyor (CLASS): 38-GHz detector array of bolometric polarimeters},'' in
  [{\em Proceedings of the SPIE, Volume 9153, id. 91531J 15 pp.
  (2014).}{\nolinebreak\hspace{0.1em}]},   {\bf 9153} (July 2014).

\bibitem{kevin}
{Denis}, K.~L., {Cao}, N.~T., {Chuss}, D.~T., {Eimer}, J., {Hinderks}, J.~R.,
  {Hsieh}, W.-T., {Moseley}, S.~H., {Stevenson}, T.~R., {Talley}, D.~J.,
  {U.-yen}, K., and {Wollack}, E.~J., ``{Fabrication of an Antenna-Coupled
  Bolometer for Cosmic Microwave Background Polarimetry},'' in [{\em American
  Institute of Physics Conference Series}{\nolinebreak\hspace{0.1em}]},
  {Young}, B., {Cabrera}, B., and {Miller}, A., eds., {\em American Institute
  of Physics Conference Series} {\bf 1185},  371--374 (Dec. 2009).

\bibitem{karwan16}
{Rostem}, K., {Ali}, A., {Appel}, J.~W., {Bennett}, C.~L., {Brown}, A.,
  {Chang}, M.-P., {Chuss}, D.~T., {Colazo}, F.~A., {Costen}, N., {Denis},
  K.~L., {Essinger-Hileman}, T., {Hu}, R., {Marriage}, T.~A., {Moseley}, S.~H.,
  {Stevenson}, T.~R., {U-Yen}, K., {Wollack}, E.~J., and {Xu}, Z.,
  ``{Silicon-based antenna-coupled polarization-sensitive millimeter-wave
  bolometer arrays for cosmic microwave background instruments},'' in [{\em
  Proceedings of the SPIE, Volume 9914, id. 99140D 10 pp.
  (2016).}{\nolinebreak\hspace{0.1em}]},   {\bf 9914} (July 2016).

\bibitem{jeff}
{Iuliano}, J., {Eimer}, J., {Parker}, L., {Ali}, A., {Appel}, J.~W., {Bennett},
  C., {Brewer}, M., {Bustos}, R., {Chuss}, D., {Cleary}, J., {Couto}, J.,
  {Dahal}, S., {Denis}, K., {D\"{u}nner}, R., {Essinger-Hileman}, T., {Fluxa},
  P., {Halpern}, M., {Harrington}, K., {Helson}, K., {Hilton}, G., {Hinshaw},
  G., {Hubmayr}, J., {Karakla}, J., {Marriage}, T., {Miller}, N., {McMahon},
  J.~J., {Nu\~{n}ez}, C., {Padilla}, I., {Palma}, G., {Petroff}, M.,
  {M\'{a}rquez}, B.~P., {Reeves}, R., {Reintsema}, C., {Rostem}, K., {Valle},
  D.~A.~N., {Van Engelhoven}, T., {Wang}, B., {Wang}, Q., {Watts}, D.,
  {Weiland}, J., {Wollack}, E.~J., {Xu}, Z., {Yan}, Z., and {Zeng}, L., ``{The
  Cosmology Large Angular Scale Surveyor Receiver Design},'' in [{\em
  Millimeter, Submillimeter, and Far-Infrared Detectors and Instrumentation for
  Astronomy IX}{\nolinebreak\hspace{0.1em}]},  {\em \procspie} {\bf 10708}
  (June 2018).

\bibitem{aamir}
Ali, A., {\em {Detectors and Focal Planes for the Cosmology Large Angular Scale
  Surveyor}}, PhD thesis, Johns Hopkins University (2017).

\bibitem{karwan14_spie}
{Rostem}, K., {Ali}, A., {Appel}, J.~W., {Bennett}, C.~L., {Chuss}, D.~T.,
  {Colazo}, F.~A., {Crowe}, E., {Denis}, K.~L., {Essinger-Hileman}, T.,
  {Marriage}, T.~A., {Moseley}, S.~H., {Stevenson}, T.~R., {Towner}, D.~W.,
  {U-Yen}, K., and {Wollack}, E.~J., ``{Scalable background-limited
  polarization-sensitive detectors for mm-wave applications},'' in [{\em
  Millimeter, Submillimeter, and Far-Infrared Detectors and Instrumentation for
  Astronomy VII}{\nolinebreak\hspace{0.1em}]},  {\em \procspie} {\bf 9153},
  91530B (July 2014).

\bibitem{zeng}
{Zeng}, L., {Bennett}, C.~L., {Chuss}, D.~T., and {Wollack}, E.~J., ``{A Low
  Cross-Polarization Smooth-Walled Horn With Improved Bandwidth},'' {\em IEEE
  Transactions on Antennas and Propagation}~{\bf 58},  1383--1387 (Apr. 2010).

\bibitem{choke}
Wollack, E.~J., U-yen, K., and Chuss, D.~T., ``Photonic choke-joints for
  dual-polarization waveguides,'' in [{\em 2010 IEEE MTT-S International
  Microwave Symposium}{\nolinebreak\hspace{0.1em}]},   177--180 (May 2010).

\bibitem{backshort}
Crowe, E.~J., Bennett, C.~L., Chuss, D.~T., Denis, K.~L., Eimer, J., Lourie,
  N., Marriage, T., Moseley, S.~H., Rostem, K., Stevenson, T.~R., Towner, D.,
  U-yen, K., and Wollack, E.~J., ``Fabrication of a silicon backshort assembly
  for waveguide-coupled superconducting detectors,'' {\em IEEE Transactions on
  Applied Superconductivity}~{\bf 23},  2500505--2500505 (June 2013).

\bibitem{chuss}
{Chuss}, D.~T., {Ali}, A., {Amiri}, M., {Appel}, J., {Bennett}, C.~L.,
  {Colazo}, F., {Denis}, K.~L., {D{\"u}nner}, R., {Essinger-Hileman}, T.,
  {Eimer}, J., {Fluxa}, P., {Gothe}, D., {Halpern}, M., {Harrington}, K.,
  {Hilton}, G., {Hinshaw}, G., {Hubmayr}, J., {Iuliano}, J., {Marriage}, T.~A.,
  {Miller}, N., {Moseley}, S.~H., {Mumby}, G., {Petroff}, M., {Reintsema}, C.,
  {Rostem}, K., {U-Yen}, K., {Watts}, D., {Wagner}, E., {Wollack}, E.~J., {Xu},
  Z., and {Zeng}, L., ``{Cosmology Large Angular Scale Surveyor (CLASS) Focal
  Plane Development},'' {\em Journal of Low Temperature Physics}~{\bf 184},
  759--764 (Aug. 2016).

\bibitem{nist}
Doriese, W.~B., Morgan, K.~M., Bennett, D.~A., Denison, E.~V., Fitzgerald,
  C.~P., Fowler, J.~W., Gard, J.~D., Hays-Wehle, J.~P., Hilton, G.~C., Irwin,
  K.~D., Joe, Y.~I., Mates, J. A.~B., O'Neil, G.~C., Reintsema, C.~D., Robbins,
  N.~O., Schmidt, D.~R., Swetz, D.~S., Tatsuno, H., Vale, L.~R., and Ullom,
  J.~N., ``Developments in time-division multiplexing of x-ray transition-edge
  sensors,'' {\em Journal of Low Temperature Physics}~{\bf 184},  389--395 (Jul
  2016).

\bibitem{nist2}
{Irwin}, K.~D., {Cho}, H.~M., {Doriese}, W.~B., {Fowler}, J.~W., {Hilton},
  G.~C., {Niemack}, M.~D., {Reintsema}, C.~D., {Schmidt}, D.~R., {Ullom},
  J.~N., and {Vale}, L.~R., ``{Advanced Code-Division Multiplexers for
  Superconducting Detector Arrays},'' {\em Journal of Low Temperature
  Physics}~{\bf 167},  588--594 (June 2012).

\bibitem{mce}
Battistelli, E.~S., Amiri, M., Burger, B., Halpern, M., Knotek, S., Ellis, M.,
  Gao, X., Kelly, D., MacIntosh, M., Irwin, K., and Reintsema, C., ``Functional
  description of read-out electronics fortime-domain multiplexed bolometers
  formillimeter and sub-millimeter astronomy,'' {\em Journal of Low Temperature
  Physics}~{\bf 151},  908--914 (May 2008).

\bibitem{karwan14}
{Rostem}, K., {Chuss}, D.~T., {Colazo}, F.~A., {Crowe}, E.~J., {Denis}, K.~L.,
  {Lourie}, N.~P., {Moseley}, S.~H., {Stevenson}, T.~R., and {Wollack}, E.~J.,
  ``{Precision control of thermal transport in cryogenic single-crystal silicon
  devices},'' {\em Journal of Applied Physics}~{\bf 115},  124508 (Mar. 2014).

\bibitem{niemack}
Niemack, M., {\em {Towards Dark Energy}}, PhD thesis, Princeton U. (2008).

\bibitem{act}
Duff, S.~M., Austermann, J., Beall, J.~A., Becker, D., Datta, R., Gallardo,
  P.~A., Henderson, S.~W., Hilton, G.~C., Ho, S.~P., Hubmayr, J., Koopman,
  B.~J., Li, D., McMahon, J., Nati, F., Niemack, M.~D., Pappas, C.~G.,
  Salatino, M., Schmitt, B.~L., Simon, S.~M., Staggs, S.~T., Stevens, J.~R.,
  Van~Lanen, J., Vavagiakis, E.~M., Ward, J.~T., and Wollack, E.~J., ``Advanced
  actpol multichroic polarimeter array fabrication process for 150 mm wafers,''
  {\em Journal of Low Temperature Physics}~{\bf 184},  634--641 (Aug 2016).

\bibitem{Irwin}
Irwin, K. and Hilton, G.,  [{\em Transition-Edge
  Sensors}{\nolinebreak\hspace{0.1em}]},  63--150, Springer Berlin Heidelberg,
  Berlin, Heidelberg (2005).

\bibitem{boyle}
Boyle, W.~S. and Rodgers, K.~F., ``Performance characteristics of a new
  low-temperature bolometer,'' {\em J. Opt. Soc. Am.}~{\bf 49},  66--69 (Jan
  1959).

\bibitem{karwan13}
{Rostem}, K., {Chuss}, D.~T., {Lourie}, N.~P., {Voellmer}, G.~M., and
  {Wollack}, E.~J., ``{A waveguide-coupled thermally isolated radiometric
  source},'' {\em Review of Scientific Instruments}~{\bf 84},
  044701--044701--6 (Apr. 2013).

\end{thebibliography}
\bibliographystyle{spiebib} % makes bibtex use spiebib.bst

\end{document}